\documentclass[aps,prb,groupedaddress,twocolumn]{revtex4-1}

\usepackage[colorlinks=true,citecolor=black,
linkcolor=black,urlcolor=blue]{hyperref}

\usepackage{bm}
\usepackage{graphics}
\usepackage{graphicx}
\usepackage{amsthm}
\usepackage{amsmath}
\usepackage{wasysym}
\usepackage{amssymb}
\usepackage[dvips]{epsfig}

\DeclareMathOperator{\tr}{tr}

\begin{document}

\title{Hall Viscosity and Momentum Transport in Lattice and Continuum Models
  of the Integer Quantum Hall Effect in Strong Magnetic Fields}

\author{Thomas I. Tuegel}
\author{Taylor L. Hughes}
\affiliation{Department of Physics and Institute for Condensed Matter Theory,\\
  University of Illinois at Urbana Champaign, IL 61801, USA}

\begin{abstract}
The Hall viscosity describes a non-dissipative response to strain in systems with broken
time-reversal symmetry. We develop a new method for computing the
Hall viscosity of lattice systems in strong magnetic fields based on momentum
transport, which we compare to the method of momentum polarization used by Tu et
al. [Phys. Rev. B \textbf{88} 195412 (2013)] and Zaletel et al. [Phys. Rev.
Lett. \textbf{110} 236801 (2013)] for non-interacting systems. We compare the
Hall viscosity of square-lattice tight-binding models in magnetic field to the
continuum integer quantum Hall effect (IQHE) showing agreement when the magnetic length is
much larger than the lattice constant, but deviation as the magnetic field
strength increases. We also relate the Hall viscosity of relativistic electrons
in magnetic field (the Dirac IQHE) to the conventional IQHE. The Hall viscosity
of the lattice Dirac model in magnetic field agrees with the continuum Dirac Hall
viscosity when the magnetic length is much larger than the lattice constant. We
also show that the Hall viscosity of the lattice model deviates further from the
continuum model if the $C_4$ symmetry of the square lattice is broken to $C_2$,
but the deviation is again minimized as the magnetic length increases.
\end{abstract}

\maketitle

\section{Introduction}
The topological response properties of the quantum Hall effect have been intensely studied for more than three decades, and  begun with the understanding that the quantized integer/fractional Hall conductance itself is a topological phenomenon.\cite{laughlin1981,tknn1982,niu1985,wenbook,fradkinbook}  The field has since understood that quantum Hall systems (with and without magnetic fields) also exhibit remarkable responses to changes in geometry.\cite{wenzee,avron_viscosity_1995,levay_berry_1995,avron1998,read_non-abelian_2009,tokatly2009,haldane_hall_2009,read_hall_2011,hughes2011,kimura_hall_2010,stone2012,bradlyn_kubo_2012,barkeshli2012,hoyos_hall_2012,wiegmann2012,ryu2012,zaletel_topological_2013,hughes2013,hidaka2013,abanov2013,biswas_semiclassical_2013,cho2014,fremling2014,bradlyn_low-energy_2014,abanov_electromagnetic_2014,parrikar2014,gromov2014b,can2014,park2014,gromov2015b,gromov2015,cho2015b} 

One interesting piece of the geometric response is the non-dissipative
Hall viscosity.\cite{avron_viscosity_1995} The Hall viscosity $\eta_H$
is an off-diagonal response coefficient that is only non-vanishing
when time-reversal symmetry is broken.
Under time-dependent shear strain, the viscosity tensor $\eta$ relates the stress tensor
$T$ to the strain rate $\dot{u}$:
\begin{equation}
  T^{\mu\nu} = -\eta^{\mu\nu\alpha\beta} \dot{u}_{\alpha\beta},
  \nonumber
\end{equation}\noindent where the strain tensor is constructed from a symmetrized gradient of the local displacement $u_{\alpha}.$
If only dissipative viscosity coefficients are present, e.g. the bulk and shear viscosities, then $\eta^{\mu\nu\alpha\beta}=\eta^{\alpha\beta\mu\nu}$ and is thus symmetric under exchange of $(\mu\nu)$ with $(\alpha\beta).$ However, being non-dissipative, the Hall viscosity generates an antisymmetric piece satisfying
\begin{equation}
  \eta_{H}^{\mu\nu\alpha\beta} = -\eta_{H}^{\alpha\beta\mu\nu}.
  \nonumber
\end{equation}
When the system is 2D and isotropic, the antisymmetric part of the
viscosity tensor is determined by a single parameter $\eta_H$ that
gives, in an isotropic 2D
orthonormal frame:~\cite{avron_viscosity_1995,avron1998}
\begin{align}
  \eta_{H}^{1112} = \eta_{H}^{1222} &= \eta_H \nonumber \\
  \eta_{H}^{1122} &= 0. \nonumber
\end{align} 

The Hall viscosity can be calculated using a variety of different methods. The first calculations were performed via the adiabatic
transport of the Hall fluid under shear strain on a torus.\cite{avron_viscosity_1995,levay_berry_1995,read_non-abelian_2009,tokatly2009} For
Schr\"odinger electrons at integer filling factors, this type of calculation yields
\begin{math}
  \eta_H = \hbar\nu\rho/4
\end{math}
where $\rho$ is the electron number
density, and $\nu$ is the integer filling fraction.~\cite{avron_viscosity_1995,levay_berry_1995,read_non-abelian_2009,read_hall_2011}
More recently, Ref. \onlinecite{bradlyn_kubo_2012} developed Kubo formulas for the
Hall viscosity which obtain the same result. Also, a new possibility for calculating the Hall viscosity was proposed via the so-called momentum polarization entanglement technique,\cite{tu_momentum_2013,zaletel_topological_2013} though there is very little explicit discussion of the results of this method in the literature (see Ref. \onlinecite{you2015} for a very recent article). Remarkably, from the adiabatic transport calculations it has been shown that for rotationally-invariant integer and fractional quantum Hall systems in large magnetic fields, the viscosity is quantized in units of the density\cite{read_non-abelian_2009} and takes the form 
\begin{align}
\eta_H=\frac{\kappa}{4}\hbar\rho
\end{align}\noindent where $\kappa$ is a universal number characterizing the particular integer/fractional quantum Hall phase, and $\rho$ is the uniform electron number density. Generically, the Hall viscosity has units of $[\tfrac{\hbar}{\ell^2}]$ for some length scale $\ell,$ but it need not always retain such a clear quantization in terms of the particle density. 

The goal of this article is two-fold: (i) we introduce a new method for the calculation of the Hall viscosity using momentum transport, and compare with the extraction of the Hall viscosity via the momentum polarization entanglement method, and (ii) we study the properties of the viscosity in two different lattice realizations of the Landau-level integer quantum Hall problem (square-lattice Hofstadter, lattice-Dirac model in a magnetic field), and illustrate the competition between contributions of the viscosity from the lattice-length scale and the magnetic-length scale.
Our article is organized as follows. In Section~\ref{sec:methods}, we present the momentum transport method
used here to compute the Hall viscosity. We also review the momentum
polarization method, which allows computation of the Hall viscosity
from the entanglement spectrum. In Section~\ref{sec:landau-level}, we
describe the application of both methods to the continuum Landau level problem,
and compare the results to previous calculations of
the Hall viscosity. In Section~\ref{sec:hofstadter}, we present
numerical calculations of the Hall viscosity for the Landau levels of a
tight-binding model (the Hofstadter
model) and discuss the results. In Section~\ref{sec:dirac-landau-level} we calculate
the Hall viscosity of the Landau levels of the continuum Dirac
equation, with and without a mass term, building on previous work
by~\textcite{kimura_hall_2010}. Finally, in Section~\ref{sec:chern}
we use a lattice analog of the continuum Dirac system in a magnetic field, and study the Hall viscosity for comparison with the continuum results.

\section{Methods}
\label{sec:methods}
We will consider two independent methods for calculating the Hall
viscosity in our example systems. The first method considers the
transverse flow of momentum when a cylinder is strained with an
area-preserving deformation. The second method uses the entanglement
spectrum to calculate the phase acquired by the many-body wavefunction
when half of a cylinder is sheared; from this phase, one can extract the
central charge\cite{tu_momentum_2013} and Hall
viscosity.\cite{zaletel_topological_2013} Let us introduce and review
both of these methods.
\paragraph{Momentum Transport}
For the first method we study the off-diagonal components of the stress tensor, which
represent the momentum flux. We will write the geometric deformation in terms of the strain tensor. To be explicit, if
$u_\alpha$ is the displacement vector, then, to lowest order, the strain
is~\cite{landau_theory_1986}
\begin{equation*}
  u_{\alpha\beta} = \frac{1}{2}\left(\partial_\alpha u_\beta
    + \partial_\beta u_\alpha\right).
\end{equation*}
In terms of the geometry, if $ds$ and $ds'$ are the original and deformed length elements, respectively,
then to lowest order in the deformations,~\cite{landau_theory_1986}
\begin{equation*}
  ds'^2 = ds^2 + 2u_{\alpha\beta} dx^\alpha dx^\beta.
\end{equation*}
Due to the structure of the Hall viscosity terms in the viscosity tensor, shear strain causes
momentum transport in the direction of momentum, but pressure/stretching
causes momentum transport orthogonal to the direction of momentum.
We find the latter is more easily studied in lattice systems when using a cylinder geometry,
therefore, we consider  deformed metrics of the form
\begin{equation}
  ds^2 = \frac{1}{\alpha^2} dx^2 + \alpha^2 dy^2
  \label{eq:cart-metric}
\end{equation}
where $\alpha$ can vary. This deformation is 
area-preserving (shear), so we need not isolate our momentum transport results from effects induced purely by changes to the density.

To calculate the momentum transport under this deformation let us consider a cylinder which is periodic in the $y$-direction with a
circumference $L_y$. As $\alpha$ varies, the strain rate is
\begin{equation}
  \dot{u} = -\frac{\dot{\alpha}}{\alpha^3}dx^2 + \alpha\dot{\alpha}\,dy^2.
  \nonumber
\end{equation}
The corresponding stress tensor components arising from the Hall viscosity contributions are
\begin{equation}
  T^{xy} = T^{yx} = -2 \eta_H \frac{\dot{\alpha}}{\alpha}.
  \nonumber
\end{equation}
Note that when the system is anisotropic ($\alpha\neq{}1$) we can have $\eta_H^{1122}\neq{}0$; unfortunately that term will not appear in this component of the stress tensor, so it cannot be extracted; but hence, it also will not affect our calculation of the other viscosity coefficients.

As the metric is deformed we want to study the amount of momentum transported from the left-half of the cylinder to the right-half. Consider cutting the cylinder at $x=x_\text{cut}$; if
$\mathcal{P}_R$ is the projection operator onto the right side of the
cut, then the total momentum on the right of the cut is
\begin{equation*}
  \left\langle P_y \mathcal{P}_R \right\rangle =
  \int_{x_\text{cut}}^\infty dx \int_0^{L_y} dy \, \Pi_y
\end{equation*}
where $\Pi_y$ is the momentum density. Typically, we will choose
$x_\text{cut}=0$ with the cylinder placed symmetrically around this
point. The stress tensor gives the momentum flux across the cut,
i.e.
\begin{equation*}
  \frac{d}{dt}\left\langle P_y\mathcal{P}_R \right\rangle =
  - \int_0^{L_y}\,dy\,T_y^{\phantom{y}x} = 2 L_y \eta_H \alpha \dot{\alpha}.
\end{equation*}
From this equation we can immediately read-off the important result:
\begin{equation}
  \eta_H =
  \frac{1}{L_y} \frac{d}{d\alpha^2}
  \left\langle{}P_y\mathcal{P}_R\right\rangle.
  \label{eq:viscosity-momentum-transport}
\end{equation} We will use this relationship between $\eta_H$ and the $\alpha$-dependence of the half-cylinder momentum to calculate the viscosity.

\paragraph{Momentum Polarization}
The second method we discuss uses the entanglement spectrum\cite{li2008,peschel2003} to determine the \emph{momentum polarization}.\cite{tu_momentum_2013} The momentum polarization was initially proposed
to calculate the topological spin and central charge of
the conformal field theory at the edge of a topological phase. For a
system in a cylindrical geometry, these data are extracted from the
expectation value of the operator $T_y^L$, which globally translates the left
half of the cylinder in the periodic direction. The expectation value
can be computed using the reduced density matrix,\cite{tu_momentum_2013}
\begin{equation}
  \lambda \equiv \left\langle{}G\right|T_y^L\left|G\right\rangle
  = \tr_L \left(\rho_{L}T_y^L\right)
\end{equation}
where $\left|G\right\rangle$ is the ground state. Ref. \onlinecite{tu_momentum_2013} shows that $\lambda$ can be easily calculated for free-fermion systems using the entanglement spectrum.

To see that the topological
spin and central charge can be extracted from this expectation value,
consider that, in the long-wavelength limit, the reduced density matrix of a cylinder cut in half
can be written in terms of the Hamiltonians $H_{Ll}$ and $H_{Lr}$ of
the respective conformal edge theories of the left and right edges of
the left half-cylinder only:\cite{tu_momentum_2013}
\begin{equation*}
  \rho_{L} = \rho_{Ll} \otimes \rho_{Lr}
  = Z^{-1} e^{-\beta_lH_{Ll} -\beta_rH_{Lr}}.
\end{equation*}
The relevant half-cylinder translation operator is
\begin{equation*}
  T_y^L = \exp{\left[\frac{2\pi{}i}{L_y}\left(P_l+P_r\right)\Delta{}y\right]},
\end{equation*}
where $\Delta{}y$ is the distance translated (which we take to be a multiple of the lattice constant for lattice systems), and $P_l$ and $P_r$ are
the generators of translations (momentum operators) of the left and
right edge theories on the half-cylinder, respectively.\cite{tu_momentum_2013,ginsparg_applied_1988}
Since the left-most edge is far from the right half,
$\beta_l\rightarrow\infty,$ and only the ground state of the left edge
contributes. The ground state expectation value of $P_l$ is $h-c/24$
where $h$ is the topological spin and $c$ is the chiral central
charge.\cite{ginsparg_applied_1988} Therefore, the contribution of
the left edge is~\cite{tu_momentum_2013}
\begin{equation*}
  \tr_{Ll}{\left(\rho_{Ll}
      \exp{\left[\frac{2\pi{}i}{L_y}P_l\Delta{}y\right]}\right)}
  = \exp{\left[\frac{2\pi{}i}{L_y}\Delta{}y
      \left(h-\frac{c}{24}\right)\right]}.
\end{equation*}
On the other hand, $\beta_r$ takes a finite value because
the right edge is entangled with the right half-cylinder.
In general, the right edge gives a non-universal
contribution~\cite{tu_momentum_2013}
\begin{equation*}
  \tr_{Lr}{\left(\rho_{Lr}
      \exp{\left[\frac{2\pi{}i}{L_y}P_r\Delta{}y\right]}\right)}
  = \exp{\left[-L_y\alpha\right]}.
\end{equation*} From this we see that one can extract the central charge and topological spin. 

For free fermions,
$\lambda$ is easily calculated in terms of the entanglement spectrum for a cylinder
by the formula~\cite{tu_momentum_2013}
\begin{equation}
  \lambda = \prod_{n,k_y}\frac{1}{2}\left[\left(1+e^{ik_y\Delta{}y}\right)
    + \left(1-e^{ik_y\Delta{}y}\right)\tanh\frac{\xi_{k_y,n}}{2}\right]
  \label{eq:momentum-polarization-ent-spectrum}
\end{equation}
where $\prod_{n,k_y}$ is a product over the bands and $y$-momenta, and
$\xi_{k_y,n}$ is the entanglement eigenvalue of the state in band $n$
with momentum $\hbar{}k_y$. The entanglement eigenvalues can be
expressed in terms of the eigenvalues of the free-electron, equal-time correlation
function,~\cite{peschel_calculation_2002}
\begin{equation}
  \xi_{k_y,n} = \log\frac{1-C^{(L)}_{k_y,n}}{C^{(L)}_{k_y,n}}
  \label{eq:free-fermion-ent-eigenvalues}
\end{equation}\noindent where $C^{(L)}_{k_y,n}$ are the eigenvalues of
$C^{(L)}_{k_y}=\langle c^{\dagger}_{k_y ia}c_{k_y jb}\rangle$ where $k_y$ are
the momenta in the periodic direction, $i,j$ run-over the lattice
sites on the left half of the cylinder, and $a,b$ run-over all of the
onsite degrees of freedom.
Note that this projects states onto the left half of the cylinder, but
we will find it more useful to compute this formula in terms of the
projections onto the right half, $C_{n,k_y}=1-C^{(L)}_{n,k_y}$.
Using these identities, it is convenient to
rewrite~\eqref{eq:momentum-polarization-ent-spectrum} as
\begin{equation}
  \lambda = \prod_{n,k_y}\frac{1}{2}\left[\left(1+e^{ik_y\Delta{}y}\right)
    + \left(1-e^{ik_y\Delta{}y}\right)\left(2C_{k_y,n}-1\right)\right].
  \label{eq:momentum-polarization-corr-func}
\end{equation}

In a remarkable extension of this work, Ref. \onlinecite{zaletel_topological_2013} shows that for quantum Hall states one can extract the Hall viscosity from the 
imaginary part of the ``non-universal" coefficient $\alpha.$ Explicitly they find 
\begin{equation}
  \lambda
  = \exp{\left[\frac{2\pi{}i}{L_y}\Delta{}y\left(h-\frac{c}{24}\right)
      - iL_y\Delta{}y\frac{\eta_H}{\hbar} + \ldots\right]}
  \label{eq:viscosity-momentum-polarization}
\end{equation}\noindent where additional non-universal terms that scale differently with $L_y$ have been dropped. In their work they consider a full twist such that $\Delta{}y=L_y,$ but the result carries over for smaller $\Delta{}y$ as well. 
 Thus, the viscosity and
central charge can be extracted from a fit of
$L_y\mathrm{Arg}\lambda$; the former from the quadratic coefficient,
the latter from the constant coefficient.

We can understand how the momentum polarization phase encodes the
viscosity by considering the action of the shear strain generators on
the ground state. Here, we will show that the Hall viscosity can be
extracted by comparing the momentum polarization calculated with a
real-space cut to the phase taken with an orbital
cut following Ref.~\onlinecite{park2014}. We note that Ref. \onlinecite{park2014} identified two distinct
contributions to the Hall viscosity, and the contribution which interests
us here is due to changing the shape of the Landau orbitals under
shear strain. The second contribution, the guiding center Hall
viscosity, comes from the electron correlations and is absent in the integer quantum Hall models we study here. We will review how the momentum polarization phase
calculated with a real-space cut encodes both Hall viscosity
contributions. Although we consider only the integer effect, the
guiding center Hall viscosity also has a super-extensive term due to
the non-zero net momentum in each half of the system.\cite{park2014}
We will show that this background can be subtracted by calculating the
momentum polarization phase with an \emph{orbital} cut and comparing the two results.

For most of the remainder of this section we closely follow Ref. \onlinecite{park2014}. First, let us decompose our physical coordinate $\mathbf{R}$ into a guiding
center coordinate $\mathbf{\widetilde{r}}$ and an orbital coordinate
$\mathbf{r}$:
\begin{equation*}
\mathbf{R} = \mathbf{\widetilde{r}} + \mathbf{r}.
\end{equation*}
There is a metric $G_{\mu\nu}$ associated with the physical coordinate
$\mathbf{R}$, as well as metrics $\widetilde{g}_{\mu\nu}$ and
$g_{\mu\nu}$ associated with each coordinate
$\mathbf{\widetilde{r}}$ and $\mathbf{r}$, respectively.
Let the operators $\widetilde{\lambda}^{\mu\nu}$ generate shear strain
(area-preserving deformations) associated with the metric
$\widetilde{g}_{\mu\nu}$; likewise, let $\lambda^{\mu\nu}$ be the
shear strain generators associated with $g_{\mu\nu}$. These generators
obey commutation relations\cite{park2014}
\begin{align}
\left[\lambda^{\mu\nu},\,\lambda^{\alpha\beta}\right]
  &= -\frac{i}{2} \left(\epsilon^{\mu\alpha}\lambda^{\nu\beta}
    + \epsilon^{\mu\beta}\lambda^{\nu\alpha}
    + \mu\leftrightarrow\nu\right) \nonumber \\
\left[\widetilde{\lambda}^{\mu\nu},\,\widetilde{\lambda}^{\alpha\beta}\right]
  &= \frac{i}{2} \left(\epsilon^{\mu\alpha}\widetilde{\lambda}^{\nu\beta}
    + \epsilon^{\mu\beta}\widetilde{\lambda}^{\nu\alpha}
    + \mu\leftrightarrow\nu\right)\nonumber\\
    \left[\widetilde{\lambda}^{\mu\nu},\,\lambda^{\alpha\beta}\right]&=0.
  \label{eq:mp-strain-commutators}
\end{align}
The strain generator in the physical coordinate is
\begin{equation*}
\Lambda^{\mu\nu} = \widetilde{\lambda}^{\mu\nu} + \lambda^{\mu\nu}
\end{equation*}
so that the unitary operator implementing strain on quantum states
is\cite{park2014,bradlyn_kubo_2012}
\begin{equation*}
U(\alpha) = \exp\left[i \int d^2\mathbf{R}\, \alpha_{\mu\nu}(\mathbf{R}) \Lambda^{\mu\nu} \right]
\end{equation*}
where $\alpha_{\mu\nu}$ is a symmetric matrix parametrizing the
strain. Because the strain generators on each coordinate commute, we
can also write this as the product of strain transformations on each coordinate:
\begin{align*}
U(\alpha) &= u(\alpha)\widetilde{u}(\alpha) \\
u(\alpha) &= \exp\left[i \int d^2\mathbf{R}\,
            \alpha_{\mu\nu}(\mathbf{R}) \lambda^{\mu\nu} \right] \\
\widetilde{u}(\alpha) &= \exp\left[i \int d^2\mathbf{R}\, \alpha_{\mu\nu}(\mathbf{R}) \widetilde{\lambda}^{\mu\nu} \right].
\end{align*}
To first order in $\alpha_{\mu\nu}$, the variation in the
metric under strain is\cite{park2014}
\begin{equation*}
  \delta G_{\mu\nu}(\mathbf{R})
  = -\epsilon^{\alpha\beta}G_{\mu\alpha}(\mathbf{R})\alpha_{\beta\nu}(\mathbf{R})
  + \mu\leftrightarrow\nu.
\end{equation*}
In our particular case, where we shear half the cylinder, this gives
\begin{equation}
  \alpha_{\mu\nu}(x,\,y) = \left(
    \begin{array}{cc}
    \delta(x)\Delta{}y & 0 \\
    0 & 0
    \end{array}\right).
  \label{eq:mp-strain}
\end{equation}
The momentum polarization expectation value $\lambda$ (c.f. 
Eq.~\eqref{eq:momentum-polarization-corr-func}) is just the ground
state expectation value $\left\langle{}U(\alpha)\right\rangle$ under
this strain field.

Before we proceed to compute the required expectation values and
find the momentum polarization phase, let us see how the Hall
viscosity enters the calculation. We can represent the viscosity tensor in terms of the 
adiabatic curvature of the ground state under shear
strain:\cite{avron_viscosity_1995,park2014}
\begin{align*}
H^{\mu\nu\alpha\beta}(\mathbf{R}) &= 2 \hbar\, \mathrm{Im}
  \left\langle\frac{d\Psi(\alpha)}{d\alpha_{\mu\nu}(\mathbf{R})} \Bigg|
  \frac{d\Psi(\alpha)}{d\alpha_{\alpha\beta}(\mathbf{R})}\right\rangle
  \\
&= -i\hbar\, \left\langle\Psi\right|
  \left[\Lambda^{\mu\nu},\,\Lambda^{\alpha\beta}\right] \left|\Psi\right\rangle.
\end{align*}
where $\left|\Psi(\alpha)\right\rangle=U(\alpha)\left|\Psi\right\rangle$.
Because the strain generator $\Lambda$ is the sum of orbital and
guiding center strain generators, we conclude that the viscosity also
has contributions due to each strain generator, which we separately
denote
\begin{align*}
  \eta^{\mu\nu\alpha\beta}(\mathbf{R})
    &= -i\hbar\, \left\langle\Psi\right|
      \left[\lambda^{\mu\nu},\,\lambda^{\alpha\beta}\right]
      \left|\Psi\right\rangle \\
  \widetilde{\eta}^{\mu\nu\alpha\beta}(\mathbf{R})
    &= -i\hbar\, \left\langle\Psi\right|
      \left[\widetilde{\lambda}^{\mu\nu},\,\widetilde{\lambda}^{\alpha\beta}\right]
      \left|\Psi\right\rangle.
\end{align*}
Now, using the strain field in Eq.~\eqref{eq:mp-strain}, we find
\begin{align*}
\lambda_{\text{RES}}
  &= \left\langle\Psi\right| U(\alpha) \left|\Psi\right\rangle \\
  &= \left\langle\Psi\right| \widetilde{u}(\alpha) \left|\Psi\right\rangle
    \left\langle\Psi\right| u(\alpha) \left|\Psi\right\rangle,
\end{align*}
where $\lambda_{\text{RES}}$ is the momentum polarization phase
$\lambda$ in Eq.~\eqref{eq:momentum-polarization-corr-func} computed with
the real-space entanglement spectrum. Now, the expectation value of
$\widetilde{u}$ is the momentum polarization phase computed
with the orbital entanglement spectrum.\cite{park2014}, while  the expectation value of $u$ is
\begin{align*}
  \left\langle\Psi\right| u(\alpha) \left|\Psi\right\rangle
  &= \left\langle\Psi\right|
    \exp\left[i \int d^2\mathbf{R}\, \alpha_{\mu\nu}(\mathbf{R}) \lambda^{\mu\nu} \right]
    \left|\Psi\right\rangle \\
  &= \left\langle\Psi\right|
    \exp\left[i \int d^2\mathbf{R}\, \alpha_{xx}(\mathbf{R}) \lambda^{xx} \right]
    \left|\Psi\right\rangle \\
  &= \left\langle\Psi\right|
    \exp\left[i \int d^2\mathbf{R}\, \delta(x) \Delta{}y \lambda^{xx} \right]
    \left|\Psi\right\rangle \\
  &= \exp\left[i \int d^2\mathbf{R}\, \delta(x) \Delta{}y
    \left\langle\Psi\right|\lambda^{xx}\left|\Psi\right\rangle \right]
  \\
  &= \exp\left[i L_y \Delta{}y
    \left\langle\Psi\right|\lambda^{xx}\left|\Psi\right\rangle \right],
\end{align*}
where we have kept terms only to first order in $\Delta{}y$. Using the
strain generator commutation relations in
Eq.~\eqref{eq:mp-strain-commutators}, we substitute
\begin{equation*}
  \left\langle\Psi\right|\lambda^{xx}\left|\Psi\right\rangle
  = i\left\langle\Psi\right|\left[\lambda^{xx},\,\lambda^{xy}\right]\left|\Psi\right\rangle
  = -\frac{1}{\hbar} \eta^{xxxy} = -\frac{1}{\hbar} \eta_H
\end{equation*}
to find
\begin{equation*}
  \left\langle\Psi\right| u(\alpha) \left|\Psi\right\rangle =
  \exp\left[-\frac{i}{\hbar} L_y \Delta{}y \eta_H \right].
\end{equation*}

Returning to our expression for the momentum polarization phase, we
have
\begin{equation*}
\lambda_{\text{RES}} = \lambda_{\text{OES}} \exp\left[-\frac{i}{\hbar}
  L_y \Delta{}y \eta_H \right],
\end{equation*}
where
\begin{math}
\lambda_{\text{OES}}=
  \left\langle\Psi\right| \widetilde{u}(\alpha)
  \left|\Psi\right\rangle
\end{math}
is the momentum polarization phase computed with the orbital
entanglement spectrum. Hence, we can determine that an alternate form of the (orbital contribution to the) Hall viscosity is given
by
\begin{equation}
\eta_H = -\frac{\hbar}{L_y \Delta{}y} \mathrm{Arg}\,
  \frac{\lambda_{\text{RES}}}{\lambda_{\text{OES}}}
  \label{eq:mp-viscosity}
\end{equation}\noindent for systems in uniform magnetic fields. 

Now that we have introduced the two separate methods for calculating the viscosity we will apply them to two different continuum systems, and their matching lattice regularized models.

\section{Continuum Landau Levels}
\label{sec:landau-level}

\begin{figure}
  \includegraphics[width=0.5\textwidth]{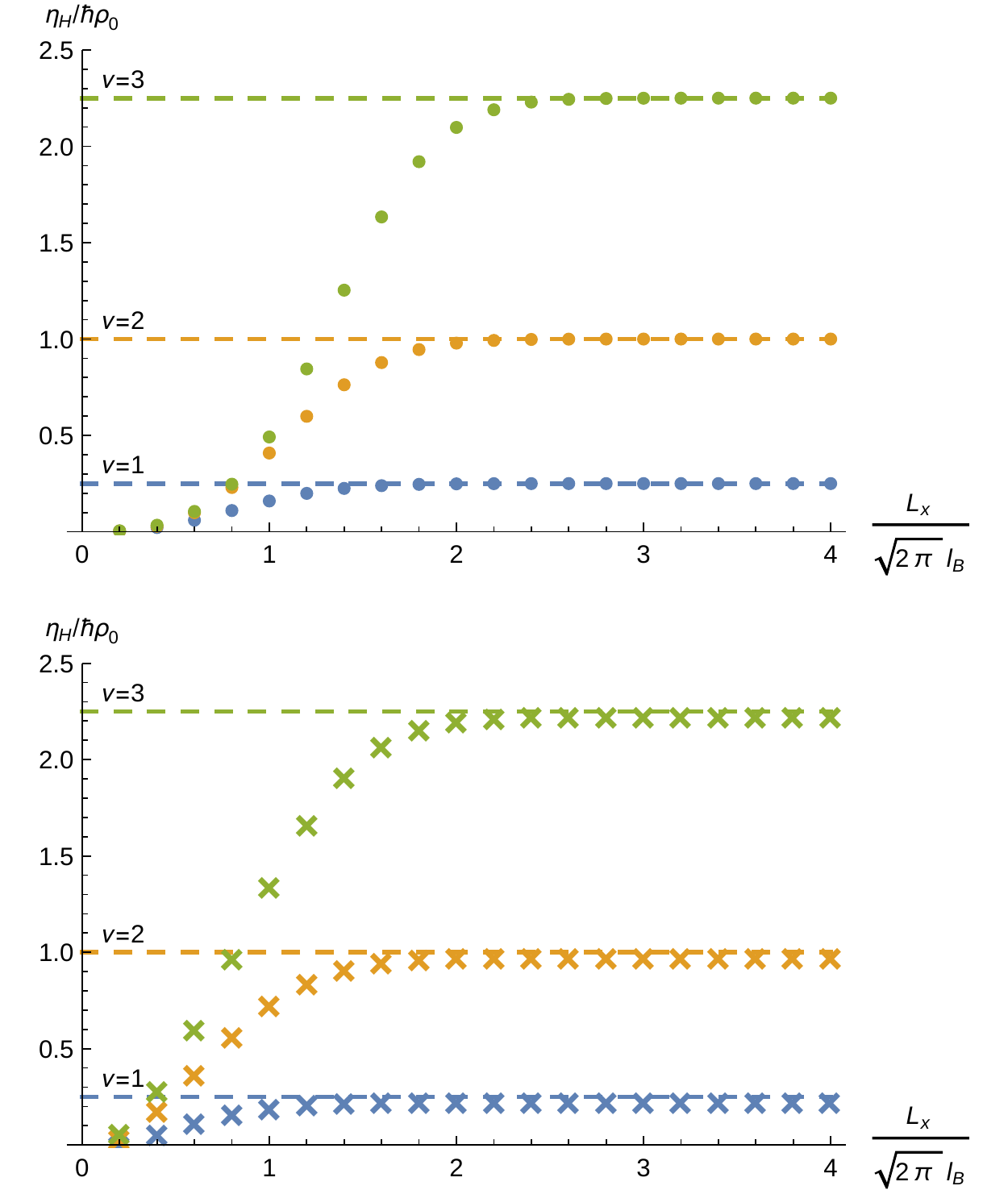}
  \caption{
    \label{fig:iqhe-convergence}
    The Hall viscosity ($\eta_H$) of the specified integer quantum
    Hall states calculated by the momentum transport (upper panel) and
    momentum polarization (lower panel) methods. The calculation converges
    when \mbox{$L_x>2\sqrt{2\pi}\ell_B$}, i.e. when each half of the
    cylinder is wider than a single wavefunction.  The Hall viscosity
    is given in units of $\hbar\rho_0$ where $\rho_0=1/2\pi{}\ell_B^2$ is
    the electron density of the lowest Landau level.
  }
\end{figure}

Let us begin with the conventional Landau level problem of 2D electrons in a uniform magnetic field, and consider the possibility of geometric deformations similar to Ref. \onlinecite{avron_viscosity_1995}. 
The Hamiltonian for electrons in a background electromagnetic field
subject to the metric of Equation~\eqref{eq:cart-metric}
is
\begin{equation*}
  H = \frac{\alpha^2}{2m}{\left(\hat{p}_x+eA_x\right)}^2 +
  \frac{1}{2m\alpha^2}{\left(\hat{p}_y+eA_y\right)}^2.
\end{equation*}
On a cylinder which is periodic in $y$ (with circumference $L_y$), and with a
uniform magnetic field normal to the cylinder (in the Landau gauge,
$A_x=0$ and $A_y=Bx$),
\begin{equation*}
  H = \frac{\alpha^2}{2m}{\hat{p}_x}^2 +
  \frac{1}{2m\alpha^2}{\left(\hat{p}_y+eB\hat{x}\right)}^2.
\end{equation*}
As is conventional, we define the lowering operator
\begin{equation}
  \hat{a}
  = \frac{1}{\sqrt{2\hbar eB}}\left[ \alpha\hat{p}_x
    - \frac{i}{\alpha}\left(\hat{p}_y+eBx\right)\right]
  \label{eq:iqhe-lowering}
\end{equation}
and its adjoint, $\hat{a}^\dag$. It is easy to verify that their
commutator is
\begin{equation*}
  \left[\hat{a},\,\hat{a}^\dag\right]=1
\end{equation*}
so that these are the usual ladder operators of quantum harmonic
oscillator. The Hamiltonian is
\begin{equation*}
  H = \hbar\omega\left(\hat{a}^\dag\hat{a} + \frac{1}{2}\right)
\end{equation*}
where $\omega=eB/m$ is the cyclotron frequency. 

The lowest Landau
level wavefunction satisfies
\begin{equation*}
  \hat{a} \, \phi_{k,\alpha}^{(0)}=0
\end{equation*}\noindent where we are using $p_y=\hbar k.$
The raising operator $\hat{a}^\dag$ generates the higher Landau levels,
\begin{equation*}
  \phi_{k,\alpha}^{(n)}
  = \frac{1}{\sqrt{n!}} \left(\hat{a}^\dag\right)^n
  \phi_{k,\alpha}^{(0)}.
\end{equation*}
The general formula for the wavefunctions for the $n$-th Landau level is
\begin{equation}
  \phi^{(n)}_{k,\alpha}(x,\,y) =
  \frac{\exp{\left[iky -\frac{{\left(x+k\ell_B^2\right)}^2}{2\alpha^2\ell_B^2}\right]}}
  {{\left(2^n n! \alpha \ell_B L_y \sqrt{\pi}\right)}^{\frac{1}{2}}}
  H_n\left(\frac{x+k\ell_B^2}{\alpha \ell_B}\right),
  \label{eq:iqhe-lll-wavefunction}
\end{equation}
with $k=2\pi{}n/L_y$ for $n\in{}\mathbb{Z},$ and where the magnetic length $\ell_{B}^2=\tfrac{\hbar}{eB}.$ $H_n$ is the $n$-th
Hermite polynomial.
When $\alpha=1$, i.e. in the absence of any metric deformation,  the wavefunctions assume their well-known isotropic form.

Let us now present the calculations for the Hall viscosity using the two methods we presented in the previous section. To calculate the Hall viscosity by the momentum transport method at a filling 
$\nu$, we need only compute the derivative (with respect to
$\alpha^2$, c.f. Eq.~\eqref{eq:viscosity-momentum-transport}) of
\begin{align}
  \left\langle{}P_y\mathcal{P}_R\right\rangle
  &= \sum_{n=0}^{\nu-1} \sum_{k=-K}^K \hbar k \, C_{k,\alpha}^{(n)}
    \label{eq:iqhe-momentum-projection} \\
  \text{where } C_{k,\alpha}^{(n)}
  &= \int_0^\infty dx \int_0^{L_y} dy
  \left|\phi_{k,\alpha}^{(n)}\left(x,\,y\right)\right|^2.
    \label{eq:iqhe-projection}
\end{align}
We note two things: (i) $C_{k,\alpha}^{(n)}$ is just the probability of finding a particle on
the right ($x>0$) half of the cylinder, given that the particle is in
the state $\phi_{k,\alpha}^{(n)},$ and (ii) these quantities match the correlation-function eigenvalues $C_{k,n}$ if one calculates the entanglement spectrum of this system by cutting the cylinder at $x=0.$ Thus the projections
$C_{k,\alpha}^{(n)}$ of the Landau level wavefunctions onto the right
half-cylinder will also used to evaluate the momentum polarization. We
list their analytic forms here for the first three Landau levels:
\newcommand{\klba}{\frac{k\ell_B}{\alpha}}
\begin{align}
  C_{k,\alpha}^{(0)}
  &= \frac{1}{2} \mathrm{erfc}\left(\klba\right)
    \nonumber \\
  C_{k,\alpha}^{(1)}
  &= \klba \frac{1}{\sqrt{\pi}}e^{-\left(k\ell_B/\alpha\right)^2}
    + \frac{1}{2} \mathrm{erfc}\left(\klba\right)
    \nonumber \\
  C_{k,\alpha}^{(2)}
  &= \left[\left(\klba\right)^3+\frac{1}{2}\klba\right]
    \frac{1}{\sqrt{\pi}}e^{-\left(k\ell_B/\alpha\right)^2}
    \nonumber \\
  &\qquad + \frac{1}{2} \mathrm{erfc}\left(\klba\right).
    \nonumber
\end{align}

Now, the range of filled $k$ states is determined by the length of the
system; the last orbital is centered at \mbox{$x=\pm K \ell_B^2$}. We find that the
viscosity derived from the sum over $k$ converges to its expected continuum value when
\mbox{$L_x>2\sqrt{2\pi}\ell_B$}, i.e. when each half of the cylinder is
wider than a single wavefunction. We show the result of the viscosity calculation when successively filling up to the first three Landau levels in 
Fig.~\ref{fig:iqhe-convergence}. We see that the
Hall viscosity contribution from each Landau level converges to the established
result~\cite{levay_berry_1995}
\begin{equation}
  \eta_H^{(n)} = \frac{\hbar}{2\pi{}\ell_B^2}
  \frac{1}{2}\left(n+\frac{1}{2}\right),
  \label{eq:iqhe-viscosity}
\end{equation} which is the Hall viscosity contribution coming from the $n$-th Landau level. 
The convergence criterion is unsurprising given our treatment of the
cylinder's edges. The edges are not sharp, rather the edge of the
cylinder is some region defined by the width of the last occupied
wavefunction. If the cylinder is narrower than the width of its edges then it is no surprise that the result does not converge properly.

Using the correlation functions above, and
Eq.~\eqref{eq:momentum-polarization-corr-func}, nearly the same results are
obtained by the momentum polarization method with
\begin{align*}
  \lambda_{\text{RES}}
  &= \prod_{n,k}\frac{1}{2}\left[\left(1+e^{ik\Delta{}y}\right)
    + \left(1-e^{ik\Delta{}y}\right)\left(2C^{(n)}_{k,\alpha}-1\right)\right]
  \\
  \lambda_{\text{OES}}
  &= \prod_{n}\prod_{k>0} e^{ik\Delta{}y} .
\end{align*}
The product over $n$ spans the occupied Landau levels. The Hall
viscosity can be calculated with Eq.~\eqref{eq:mp-viscosity} by
computing $\lambda_{\text{RES}}$ and $\lambda_{\text{OES}}$ at several
values of $L_y$ and extracting the quadratic fit coefficient. The
result is shown in the lower panel of Fig.~\ref{fig:iqhe-convergence}, which shows that
the calculation converges when \mbox{$L_x>2\sqrt{2\pi}\ell_B$}. This is
the same criterion as for the convergence of the momentum transport
calculation: each half of the cylinder must be wider than a single
wavefunction.

\section{Hofstadter Model}
\label{sec:hofstadter}
\begin{figure}
  \includegraphics{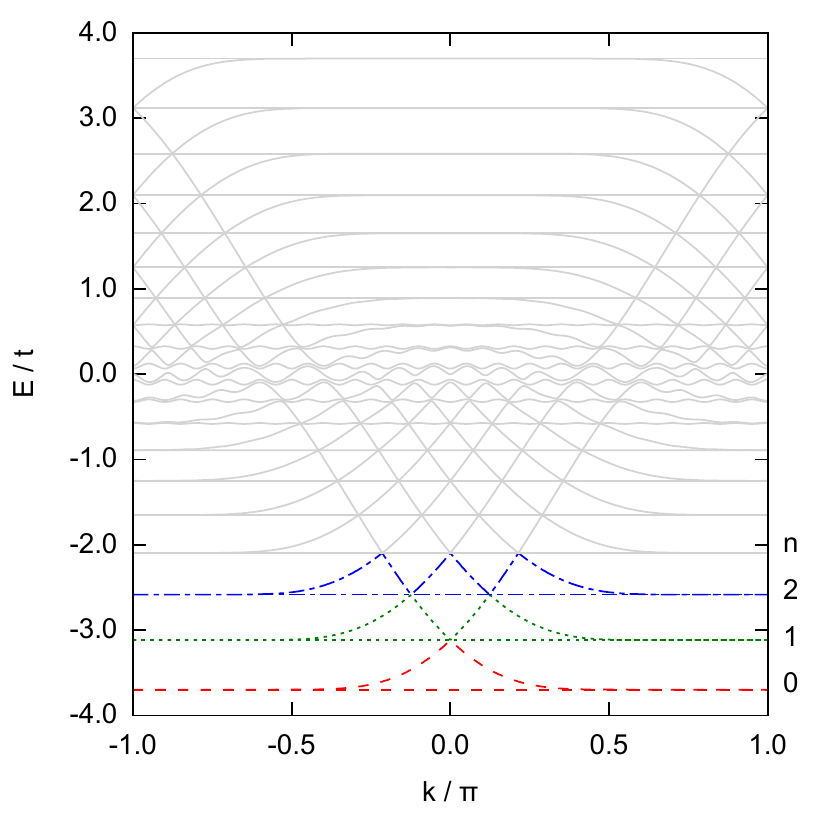}
  \caption{
    \label{fig:hof-spectrum}
    The spectrum of the Hofstadter Hamiltonian in
    Eq.~\eqref{eq:hof-hamiltonian} with $q=20$. The first three
    Landau levels are highlighted to illustrate the level filling
    scheme.
  }
\end{figure}
The utility of the two methods we have introduced is that they can be easily adapted to calculate the Hall viscosity in discrete, lattice systems in magnetic fields as well. The lattice systems have discrete translation and rotation symmetries, and have an additional length scale $a,$ the lattice constant. Since there is not continuous rotation symmetry, then we can no longer appeal to the result that the viscosity is quantized in terms of the density.\cite{read_non-abelian_2009} Furthermore, when considering momentum transport, we must consider the fact that continuous translation symmetry is broken, and thus we are really considering the transport of quasi-momentum. Additionally, for the momentum polarization technique, there is now a minimal $\Delta{}y$, i.e. the lattice constant in the $y$-direction. In lattice systems we thus might expect that there is a maximum viscosity bound that is physically meaningful, i.e. when twisting the lattice by a single-lattice constant causes the transport of a full reciprocal lattice vector of momentum, then it is as if nothing has been transported. We will save a careful discussion of some of these issues to future work. For now we will compare the results of the two methods to see if they give matching results for the lattice viscosities, and moreover, if they both converge to the continuum limit when the magnetic length becomes much longer than the lattice scale. As an aside, we note that Ref. \onlinecite{biswas_semiclassical_2013} has also performed some viscosity calculations for the Hofstadter problem in a different context/methodology and recovers the continuum limit of the viscosity for small magnetic fields.

We begin with the Hofstadter model,\cite{hofstadter1976} which is the tight-binding version of the integer quantum Hall
problem. The square lattice tight-binding model with rational flux
$\phi=p/q$ per plaquette has a Hamiltonian
\begin{equation*}
  H = \sum_{n,\,k_y} -t_x c_{n+1,k_y}^\dag c_{n,k_y} -t_y
  \cos{\left(k_y-2\pi\phi n\right)} c_{n,k_y}^\dag c_{n,k_y}
  + \text{h.c.}
\end{equation*}
on a cylinder, where $c_{n,k_y}$ annihilates an electron in the
$y$-momentum mode with wavenumber $k_y$ on the $n$-th site in the
$x$-direction. 

Although the Hofstadter model in the Landau gauge does not retain the fundamental translation symmetry of the lattice in the $x$ direction, it is symmetric
under translation by a whole magnetic cell ($q$ unit cells). To have periodic boundary conditions in a
torus geometry, we must respect this symmetry by having an integer number of
magnetic cells, $N_x=lq$ for integer $l$. Constructing the cylinder
from the properly periodic torus--by setting the wavefunction to zero
on all the sites at one $x$ coordinate--preserves the symmetry. Thus, the
cylinder has the same number of unit cells as the torus, $lq$, but it
has one fewer site, $N_x=lq-1$.~\cite{hatsugai_edge_1993} It is important to note that the consideration of the magnetic translation symmetry
 not only affects the construction of the lattice, but
also the correct scaling of the viscosity (by the lowest Landau level
density). We also recall that with commensurate boundary conditions, the spectrum of
the Hofstadter Hamiltonian on a cylinder has $q$ nearly flat bands (Landau levels)
consisting of $l-1$ states for each momentum mode $k_y$. In the gaps
between bands are edge states (one per mode $k_y$) which connect the
flat bands. We show an example of the energy spectrum with open boundary conditions and $q=20$ in Fig. \ref{fig:hof-spectrum}.

\begin{figure}
  \includegraphics{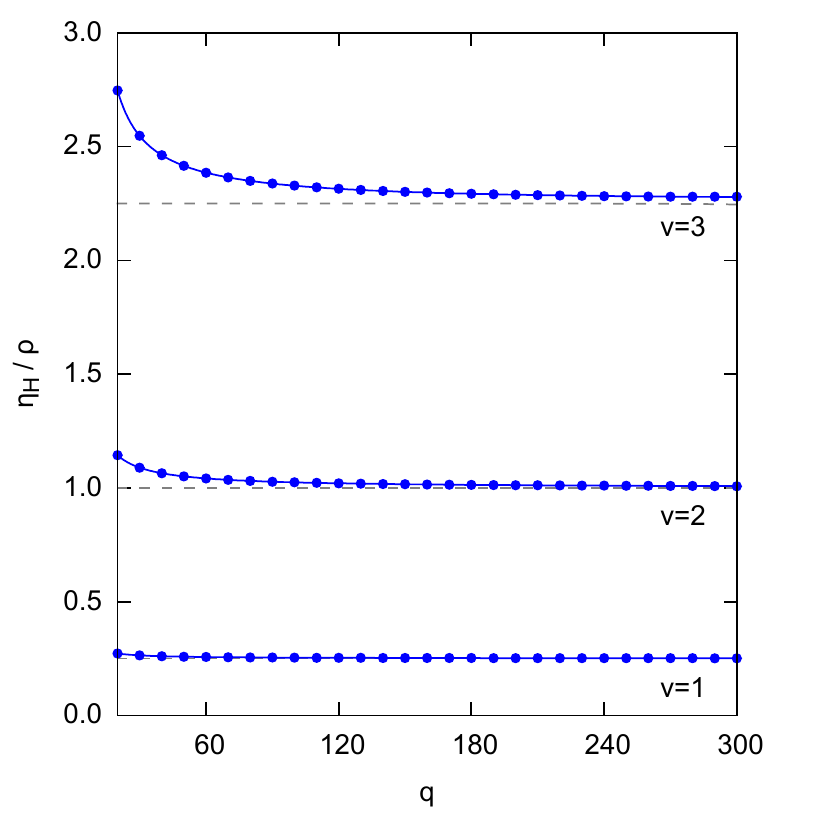}
  \caption{\label{fig:hof-viscosity-transport}
    The Hall viscosity of the Hofstadter model calculated by
    Eq.~\eqref{eq:viscosity-momentum-transport}, with $N_y=51$ and
    $N_x=2q-1$. The Hall viscosity of the continuum model is shown by a
    dashed line for comparison.
  }
\end{figure}

Since we are only deforming the diagonal components of the metric, we can the standard nearest-neighbor tight-binding model given above. Hence, the lattice is deformed through the hopping parameters $t_x$ and
$t_y$. Absent the magnetic field, the tight-binding model bands are
well-known:
\begin{equation*}
  \epsilon(k_x,\,k_y) = -2t_x\cos{(k_xa_x)} - 2t_y\cos{(k_ya_y)}.
\end{equation*}
To see how the metric will enter, we can match parameters to the continuum Hamiltonian through a long wavelength expansion
around the bottom of these bands:
\begin{equation*}
  \epsilon(k_x,\,k_y) = \sum_i \left(-2t_i + \frac{t_i}{2}{\left(k_i a_i\right)}^2\right).
\end{equation*}
We can compare this to the Schr\"odinger equation:
\begin{equation*}
  \epsilon = \epsilon_0 + \sum_{i,j} \frac{\hbar^2}{2m}k_i g^{ij}k_{j}
\end{equation*}
and then equate the coefficients of $k_i.$ Thus, we find the hopping amplitude in
each direction is inversely proportional to the lattice constant:
\begin{equation*}
  t_i = \frac{\hbar^2}{2ma_i^2}.
\end{equation*}
Therefore, under the metric deformation in Eq.~\eqref{eq:cart-metric}, we have
$t_x\propto\alpha^2$ and $t_y\propto\alpha^{-2}$. 
Hence, we consider a deformed 
Hamiltonian:
\begin{align}
  H = \sum_{n,\,k_y} &\left(-t_x\alpha^2 c_{n+1,k_y}^\dag c_{n,k_y} \right.\nonumber \\
  &\quad-\left.\frac{t_y}{\alpha^2}
  \cos{\left(k_y-2\pi\phi n\right)} c_{n,k_y}^\dag c_{n,k_y}
  + \text{h.c.}\right).
  \label{eq:hof-hamiltonian}
\end{align}

Now let us explicitly detail how the momentum-transport is calculated. The projected  \mbox{(quasi-)momentum} is
\begin{equation}
  \left\langle{}P_y\mathcal{P}_R\right\rangle
  = \sum_{k_y}\sum_{m=1}^{\nu l} \hbar k_y
  \left\langle m,\,k_y \right|\mathcal{P}_R\left| m,\,k_y \right\rangle
  \label{eq:hof-projected-momentum}
\end{equation}
where $\mathcal{P}_R$ projects onto the right half of the cylinder: 
\begin{equation*}
  \mathcal{P}_R = \sum_{x=0}^{N_x/2}
  \left|x\right\rangle\left\langle{}x\right|.
\end{equation*}
The integers $m$ run over energy eigenstates at a given $k_y$ from $1$ (lowest energy) to a value depending on the filling.  As a reminder, we point out that if the $\nu>q/2$, the edge states associated with each Landau level above the middle of the spectrum are actually below the flat Landau level, rather than above, so one would need to be careful when choosing which states are filled if the viscosity of those Landau levels is of interest.
 The filling scheme for the first few Landau levels is illustrated in
Fig.~\ref{fig:hof-spectrum}.
We also note that, for our calculations, the site at which the cylinder is  cut should fall on the
boundary between magnetic cells, i.e. on a site $n=rq$ for
$r\in{}\mathbb{Z}$. We need to impose this condition so that the subsystems
have commensurate boundary conditions, thus ensuring that the edge states
of each half cylinder are the same as the physical edge states of the whole
system.~\cite{hatsugai_edge_1993}

%TLH
The viscosity can be computed directly from the
eigenstates of the Hamiltonian~\eqref{eq:hof-hamiltonian} using
Eqs.~\eqref{eq:viscosity-momentum-transport}
and~\eqref{eq:hof-projected-momentum}. At large $q$, the
magnetic field is weak, and the magnetic length is much larger than
the lattice spacing. In this regime we expect that the Hall viscosity should approach the
continuum model. In fact, as in Fig.~\ref{fig:hof-viscosity-transport},
we see that it does converge to the continuum result for the fillings
we tested. As one increases the magnetic field, the effects of the
lattice will become more prominent. This figure also indicates that
lattice effects more strongly affect higher Landau levels since the
convergence to the continuum limit is slower. Eventually, as the magnetic field strengthens, i.e. as $q\to 0$, the
viscosity begins to depend on the lattice scale. From our results in the continuum we expect that, when divided by the density, the viscosity should be a constant, independent of $q.$ Instead we find that the viscosity  has contributions that depend on $q$:
\begin{align*}
\frac{2\pi\ell_{B}^2}{\hbar} \eta_H^{(1)}
  &\sim 0.2499 + \frac{0.0017}{\sqrt{q}} + \frac{0.3865}{q} \\
  &= 0.2499 + \frac{0.0017}{\sqrt{2\pi}}\frac{a}{\ell_B} + \frac{0.3865}{2\pi}\frac{a^2}{\ell_B^2} \\
\frac{2\pi\ell_{B}^2}{\hbar} \eta_H^{(2)}
  &\sim 1.0042 - \frac{0.1513}{\sqrt{q}} + \frac{4.3204}{q} \\
  &= 1.0042 + \frac{0.1513}{\sqrt{2\pi}}\frac{a}{\ell_B} + \frac{4.3204}{2\pi}\frac{a^2}{\ell_B^2} \\
\frac{2\pi\ell_{B}^2}{\hbar} \eta_H^{(3)}
  &\sim 2.2289 - \frac{0.5938}{\sqrt{q}} + \frac{2.2256}{q} \\
  &= 2.2289 + \frac{0.5938}{\sqrt{2\pi}}\frac{a}{\ell_B} + \frac{2.2256}{2\pi}\frac{a^2}{\ell_B^2},
\end{align*}
where we have rewritten the $q$ dependence in terms of the relevant length scales using the fact that $qa^2=2\pi{}\ell_B^2$. We find that the viscosity is unchanged under $B\rightarrow{}-B$, including the $q$-dependent contributions.

\begin{figure}
  \includegraphics{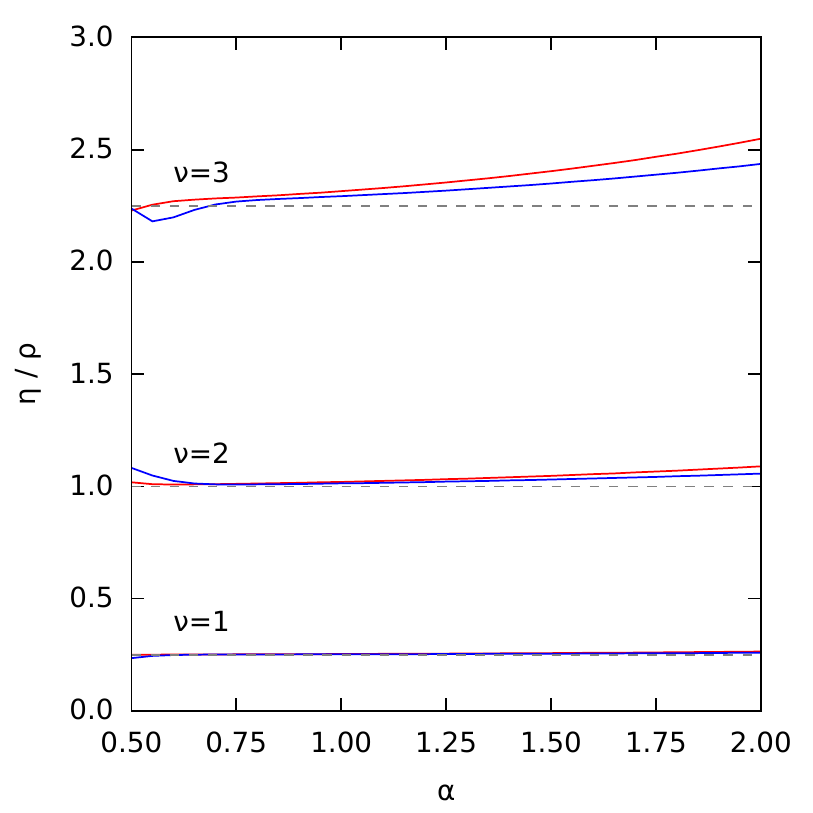}
  \caption{\label{fig:hof-viscosity-aniso}
    The Hall viscosity of the Hofstadter model at $q=120$ (red) and
    $q=180$ (blue) as a function of $\alpha$ in comparison to the
    continuum quantum Hall model (dashed gray) for filling
    $\nu=1,2,3$. The system is isotropic when $\alpha=1$ and the unit
    cells are elongated in the $y$ direction when $\alpha>1$.
  }
\end{figure}

The calculations in Fig.~\ref{fig:hof-viscosity-transport} are performed using derivatives with respect to the metric deformation $\alpha$, but evaluated at the isotropic point $\alpha=1.$ For comparison, in Fig.~\ref{fig:hof-viscosity-aniso} we fix $q$ large enough $(q=120, 180)$ so that the viscosities for the first three Landau levels have (nearly) saturated at the continuum limit, and then evaluate the momentum transport at different values of $\alpha.$ That is, we see how deforming around an initially anisotropic system affects the calculation. For values of $\alpha\neq 1$ the system only has $180^{\circ}$-rotation symmetry, and is quasi-1D for large deviations. Fig.~\ref{fig:hof-viscosity-aniso} shows that the
viscosity of the Hofstadter model varies as a function of $\alpha$
itself; in comparison, the viscosity of the continuum Landau level is
constant as $\alpha$ is varied. If
the system is anisotropic we would expect the Hall viscosity to be
controlled by more than one coefficient, for example
$\eta_{H}^{1112}\neq\eta_{H}^{1222}$ or $\eta_H^{1122}\neq{}0.$
Helpfully, because of the $\alpha\to \alpha^{-1}$ symmetry of the
metric when we switch $x$ and $y,$ we should be able to read off both
viscosity coefficients from the same figure if we consider both
$\alpha$ and $\alpha^{-1}$ simultaneously. However, we do not expect
$\eta_H^{1122}$ to enter the momentum transport calculation, and so we
cannot extract that coefficient from this figure. Finally, we note
that if we directly compare the results at $q=120$ and $q=180$ as
shown in Fig.~\ref{fig:hof-viscosity-aniso}, we find that dependence
of the Hall viscosity on the anisotropy is weaker for larger $q$,
which might be expected since lattice effects will naturally be less
important when $\ell_B\gg a.$ In future work, it would be interesting
to see if any remnants of lattice anisotropy might survive to affect
the thermodynamic limit.

\begin{figure}
  \includegraphics{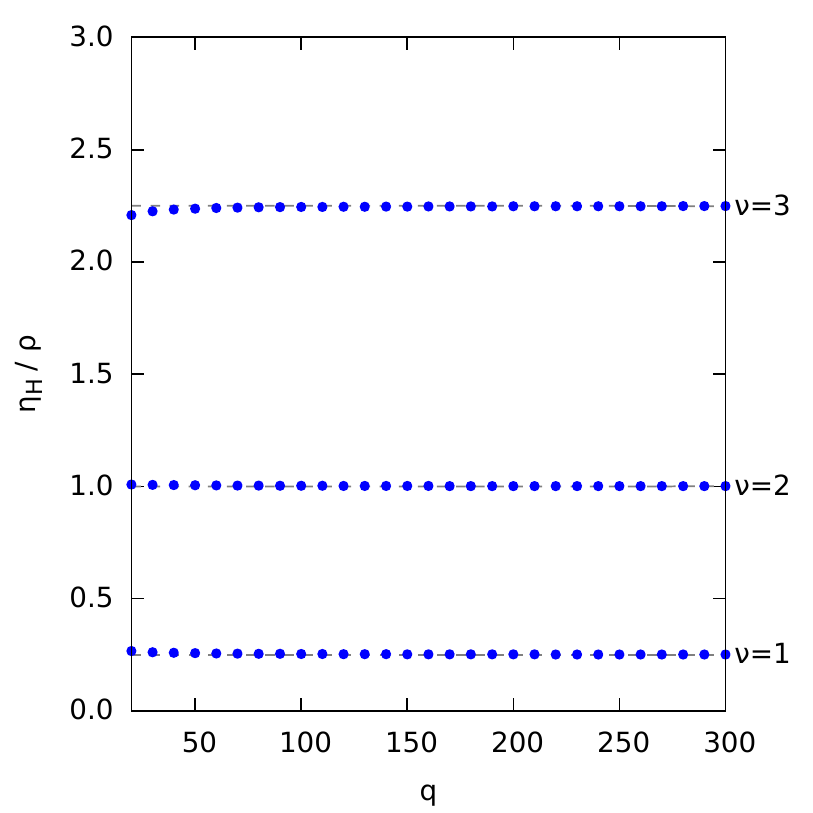}
  \caption{\label{fig:hof-polar-viscosity}
    The Hall viscosity of the Hofstadter model calculated by
    Eqs.~\eqref{eq:mp-viscosity},
\eqref{eq:hof-polar}, with $N_x=2q-1$. The Hall viscosity of
    the continuum model is shown by a dashed line for comparison.
  }
\end{figure}

To close this section, let us compare these results with those obtained from the momentum polarization
method. Note that in Eq.~\eqref{eq:hof-projected-momentum}, the
factor
\begin{equation*}
  C^{(m)}_{k_y,\alpha}
  = \left\langle m,\,k_y \right|\mathcal{P}_R\left| m,\,k_y \right\rangle
\end{equation*}
is just the aforementioned correlation function. From
Eq.~\eqref{eq:momentum-polarization-corr-func}, we can use the
correlation function to compute the momentum polarization phases:
\begin{align}
  \lambda_{\text{RES}}
  &= \prod_{m,k_y} \frac{1}{2}
    \left[\left(1+e^{ik_y\Delta{}y}\right)\right.
    \nonumber \\
  &\qquad + \left.\left(1-e^{ik_y\Delta{}y}\right)
    \left(2\left\langle m,\,k_y \right|\mathcal{P}_R\left|
        m,\,k_y \right\rangle-1\right)\right]
    \nonumber \\
  \lambda_{\text{OES}}
  &= \prod_{m,k_y} \frac{1}{2}
    \left[\left(1+e^{ik_y\Delta{}y}\right)\right.
    \nonumber \\
  &\qquad + \left.\left(1-e^{ik_y\Delta{}y}\right)
    \left(2\Theta(\left\langle m,\,k_y \right|\hat{x}\left|
        m,\,k_y \right\rangle)-1\right)\right],
    \label{eq:hof-polar}
\end{align}
where $\Theta$ is the Heaviside step function, and $\hat{x}$ is the
$x$-coordinate operator.  On a lattice, $\Delta{}y$ must be an integer
in units of the lattice constant. The resulting viscosity calculation
is shown in Fig.~\ref{fig:hof-polar-viscosity}. The Hall viscosity
obtained for the first three Landau levels agrees with the continuum
value in the weak field limit. Where the calculation converges,
i.e. $q\gtrsim 20,$ it agrees qualitatively with the momentum
transport method, although the momentum polarization calculation
appears to deviate less from the continuum Hall viscosity at small
$q$.

\section{Continuum Dirac Landau Levels}
\label{sec:dirac-landau-level}
For our second set of examples we focus on 2D Dirac fermions in a magnetic field. The quantum Hall effect in this type of system became fundamentally important with the rise of graphene,\cite{neto2009} and more recently has become relevant in the study of 3D topological insulators with low-energy surface fermions of Dirac nature.\cite{hasan2010}
We can describe the Landau level problem in the (massive) Dirac Hamiltonian under shear strain with the Hamiltonian
\begin{equation}
  H = \alpha\left(\hat{p}_x+eA_x\right)\sigma^x
  + \alpha^{-1}\left(\hat{p}_y+eA_y\right)\sigma^y
  + m \sigma^z
  \label{eq:dirac-hamiltonian}
\end{equation}
where we have again chosen the $\alpha$-dependent metric/frame in Eq.~\eqref{eq:cart-metric}, and  $\sigma^a$ are the usual Pauli matrices. As above for the
Schr\"odinger equation, let us consider a cylinder which is periodic in the $y$-direction with circumference $L_y,$ and in the Landau gauge where $A_x=0$
and $A_y=Bx$ for a uniform magnetic field normal to the surface. Again
the Hamiltonian can be written in terms of the raising and lowering
operators~\eqref{eq:iqhe-lowering}:
\begin{equation*}
  H = \left(
    \begin{array}{cc}m & \sqrt{2\hbar{}eB}\hat{a} \\
      \sqrt{2\hbar{}eB}\hat{a}^\dag & -m\end{array}
  \right).
\end{equation*}
The Landau level wavefunctions are, up to normalization,
\begin{align}
  \psi^{(0)}
  & = \left(\begin{array}{c} 0 \\ \phi_{k,\alpha}^{(0)}\end{array}\right)
  \nonumber
  \\
  \psi^{\pm(n)}
  &= \frac{1}{\sqrt{n+{p_{\pm n}(\gamma)}^2}}
    \left(\begin{array}{c}
            p_{\pm n}(\gamma) \phi_{k,\alpha}^{(n)}
            \\
            \sqrt{n} \phi_{k,\alpha}^{(n+1)}\end{array}\right)
  \nonumber
\end{align}
where $\gamma=m/\sqrt{2\hbar{}eB}$ is the ratio of the two energy
scales in the problem, $\phi_{k,\alpha}^{(n)}$ are the Schr\"odinger Landau level
wavefunctions wavefunctions~\eqref{eq:iqhe-lll-wavefunction}, $p_y=\hbar k,$ and we
have denoted for convenience $p_{\pm n}(\gamma)=\gamma\pm\sqrt{\gamma^2+n}$.
The energies of each Landau level are
\begin{align}
  E_0 & = -\gamma \sqrt{2\hbar{}eB} =-m\nonumber \\
  E_{\pm n} &= \pm \sqrt{2\hbar{}eB}\sqrt{\gamma^2+n}. \nonumber
\end{align}

Now that we have the Landau-level wavefunctions, we can  calculate the viscosity
using Eq.~\eqref{eq:viscosity-momentum-transport}, i.e. by
differentiating
\begin{align}
  \left\langle{}P_y\mathcal{P}_R\right\rangle^{(0)}
  &= \sum_{k=-K}^K \hbar k \, C_{k,\alpha}^{(0)} \nonumber \\
  \left\langle{}P_y\mathcal{P}_R\right\rangle^{(n \neq 0)}
  &= \sum_{k=-K}^K \hbar k \,
    \frac{nC_{k,\alpha}^{(n)}+{p_{\pm n}(\gamma)}^2C_{k,\alpha}^{(n-1)}}{n+{p_{\pm n}(\gamma)}^2}
  \label{eq:iqhe-dirac-momentum-projection}
\end{align}
with respect to $\alpha^2$.  Recall that $C_{k,\alpha}^{(n)}$ is defined in
Eq.~\eqref{eq:iqhe-projection}, and matches our earlier results since the Dirac Landau-levels are constructed from the Schr\"odinger Landau-levels.

\begin{figure}
  \includegraphics[width=0.5\textwidth]{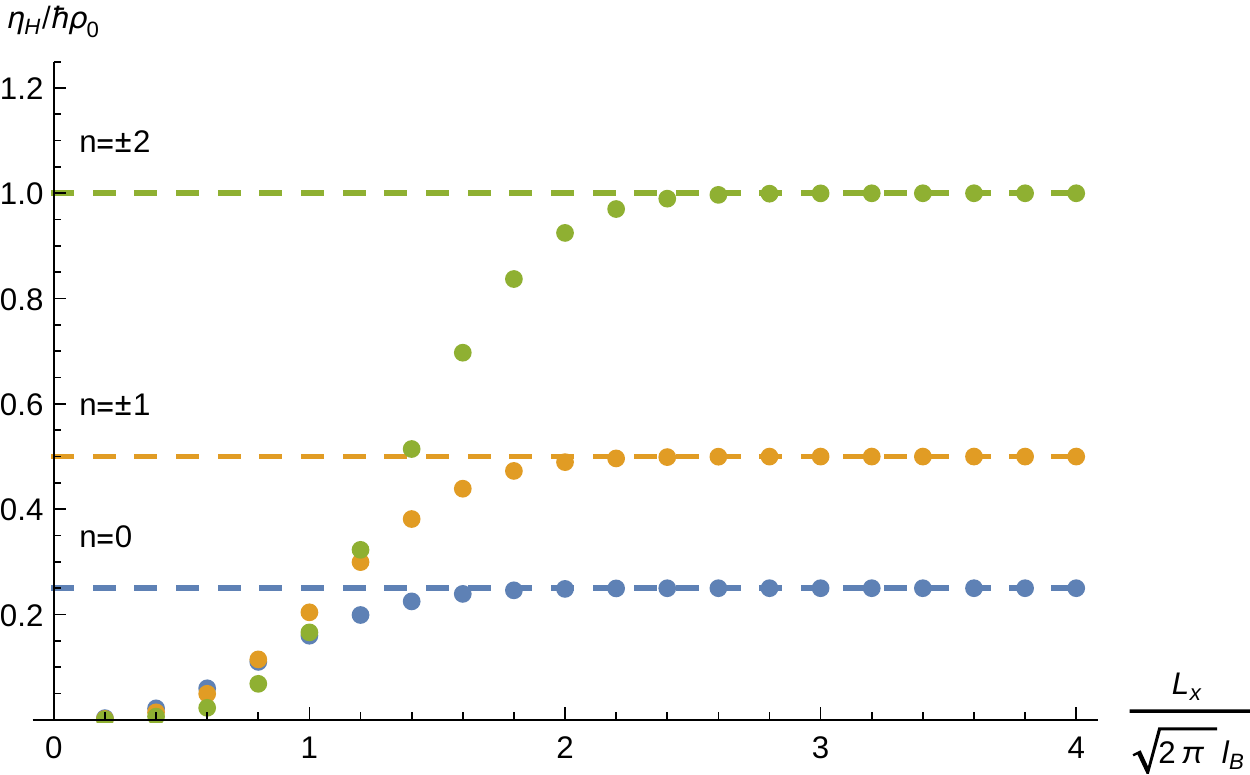}
  \caption{
    \label{fig:iqhe-dirac-transport-convergence}
    The Hall viscosity ($\eta_H$) of the specified integer quantum
    Hall states of the Dirac model~\eqref{eq:dirac-hamiltonian}
    calculated by the momentum transport method. As in
    Fig.~\ref{fig:iqhe-convergence}, the calculation converges when
    \mbox{$L_x>2\sqrt{2\pi}\ell_B$}. The derivatives in
    Eq.~\eqref{eq:viscosity-momentum-transport} were taken numerically
    with $\left\langle{}P_y\mathcal{P}_R\right\rangle$ given by
    Eq.~\eqref{eq:iqhe-dirac-momentum-projection}. The Hall viscosity
    is given in units of $\hbar\rho_0$ where $\rho_0=1/2\pi{}\ell_B^2$ is
    the electron density of the lowest Landau level.
  }
\end{figure}

Because of the connection between the Dirac and Schr\"odinger Landau-levels we conclude that the Hall
viscosity of each Landau level in the continuum Dirac system is given by
\begin{equation}
  \eta_{H,D}^{(n)} =
  \begin{cases}
    \eta_{H,S}^{(0)}
    & n = 0
    \\ & \\
   \frac{ n\eta_{H,S}^{(n)}+{p_{\pm n}(\gamma)}^2\eta_{H,S}^{(n-1)}}
    {n+{p_{\pm n}(\gamma)}^2}
    & n \neq 0
  \end{cases}
  \label{eq:iqhe-dirac-viscosity}
\end{equation}
where $\eta_{H,S}^{(n)}$ is the Hall viscosity of the $n$th Landau
level of the continuum Schr\"odinger equation  as given in 
Eq.~\eqref{eq:iqhe-viscosity}. In the massless limit, when $m=\gamma=0$ so that $p_{\pm n}=\pm\sqrt{n},$ we find
\begin{equation}
  \eta_{H,D}^{(n)} =
  \begin{cases}
    \hbar / \left(8\pi \ell_B^2\right)
    & n = 0
    \\
    \hbar \left|n\right| / \left(4\pi \ell_B^2\right)
    & n \neq 0.
  \end{cases}
\end{equation}
This result is in agreement with previous work by
\textcite{kimura_hall_2010} based on an adiabatic curvature
calculation, except for the $n=0$ level for which we have found a value
twice as large. We attribute the difference to a probable error in the
normalization of the zeroth Landau level in
Ref. \onlinecite{kimura_hall_2010}.  We confirm the results
numerically in Fig.~\ref{fig:iqhe-dirac-transport-convergence} using
the momentum transport method of calculating the Hall
viscosity. Because the Hall viscosity of the Dirac Landau levels is
expressed in terms of the Hall viscosity of the Schr\"odinger Landau
levels, the convergence criterion is expected to be the same. Indeed,
we find the result converges to the expected value when
\mbox{$L_x>2\sqrt{2\pi}\ell_B$}. The same result is obtained by the
momentum polarization method, with similar convergence criteria, though we do not show the figure here.

Let us now test if the Dirac calculation reproduces the Schr\"odinger result in the large mass limit. Thus, we will consider the $\gamma\rightarrow\pm\infty$ limit. In either limit,
\begin{align}
  p_{+n}(\gamma) &\approx 2\gamma + \frac{n}{2\gamma^2} \nonumber \\
  p_{-n}(\gamma) &\approx \frac{n}{2\gamma^2} \nonumber
\end{align}
and the resulting wavefunctions are
\begin{align}
  \psi^{+(n)}
  &\approx \left(\begin{array}{c}0 \\ \phi^{(n+1)}_{k,\alpha}\end{array}\right)
  \nonumber \\
  \psi^{-(n)}
  &\approx \left(\begin{array}{c}\phi^{(n)}_{k,\alpha} \\ 0\end{array}\right).
  \nonumber
\end{align}
The limiting values of the wavefunctions can easily be determined by considering the order, with respect
to $\gamma,$ of each component of the spinors. Additionally, the $\psi^{(0)}$ wavefunction
is completely unmodified in this limit. From this result we can conclude immediately that, in the infinite
mass limit, the Dirac Landau levels carry the same set of values of the viscosity  
as Schr\"odinger Landau levels, though we still need to see how they are organized. Additionally, in this limit the energy eigenvalues are
\begin{align}
  E_0 &= -m \nonumber \\
  E_{\pm n}
      &\approx \pm\sqrt{2\hbar eB}\left(\gamma+\frac{n}{2\gamma}\right)
      = \pm |E_0| \pm\hbar\omega n \nonumber
\end{align}
with $\omega=\left|eB/m\right|$ the usual cyclotron frequency.
The spectrum has a gap of width $2|E_0|$ with Landau levels above and
below separated from neighboring Landau levels by a gaps of uniform width $\hbar\omega$, much like
the Schr\"odinger spectrum.

The conclusions so far hold generically in the $\gamma\rightarrow\pm\infty$ limits. Let us now
consider each limit independently, and furthermore, let us consider
taking each limit by fixing $B$ and sending $m\rightarrow\pm\infty$,
respectively. In either case, the wavefunction of the $n=0$ Landau
level is essentially unchanged from the Schr\"odinger system. When
$m\rightarrow\infty$, the $n=0$ Landau level sits at the top of the
valence ($E<0$) band, separated from the $n>0$ Landau levels by the
(large) mass gap. On the other hand, when $m\rightarrow-\infty$, the $n=0$
band sits at the bottom of the conduction ($E>0$) band with only the
cyclotron gap separating it from the $n>0$ states. It is this
configuration, when $m\rightarrow-\infty,$ and with the $E<0$ states
filled, which more precisely matches the Schr\"odinger case. This should not be
surprising; the $m\sigma^z$ term of the Hamiltonian attaches a
positive mass to the $n=0$ Landau level when $m<0$. Thus, we see that the massive Dirac case matches the Schrodinger case if one focuses on the positive energy levels when $m\to -\infty.$ 

Now that we have discussed some properties of the continuum Dirac model, let us consider a lattice version.

%\begin{figure*}[h!]
%  \includegraphics{chern-spectrum-zoom.pdf}
%  \caption{
%    \label{fig:chern-spectrum-zoom}
%    A detailed view of the spectrum of the lattice Dirac Hamiltonian in
%    Eq.~\eqref{eq:chern-hamiltonian} at $m=0$ (left) and $m=4$
%    (right) showing the $n=0,\pm 1,\pm 2$ Landau levels.
%  }
%\end{figure*}

\section{Lattice Dirac Model}
\label{sec:chern}

\begin{figure}[t!]
  \includegraphics[width= .5\textwidth]{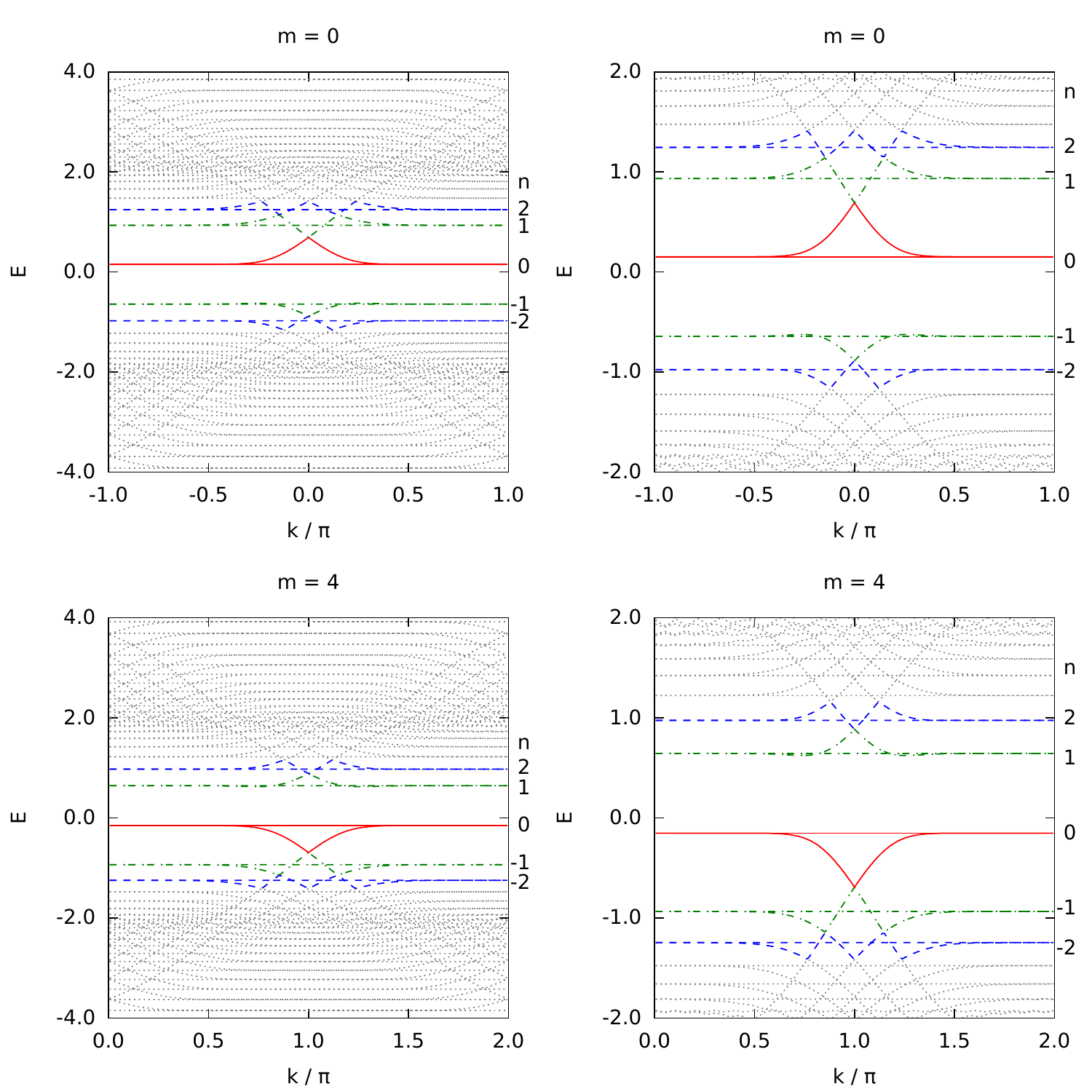}
  \caption{
    \label{fig:chern-spectrum}
    The spectrum of the lattice Dirac Hamiltonian in
    Eq.~\eqref{eq:chern-hamiltonian} at $m=0$ (top) and $m=4$ (bottom)
 The $n=0,\pm 1,\pm 2$ Landau levels are indicated. A zoomed-in 
    view of these Landau levels is presented in the right panels of each row.
    }
\end{figure}

Let us consider a lattice regularization of the continuum Dirac model. Despite the fact that the Dirac Landau-level spectrum is celebrated because of its application in graphene, we will not consider such a honeycomb lattice model. The reason is that they honeycomb model presents extra difficulties. For example,  there are not only multiple Dirac cones, but the cones are located away from the $\Gamma$-point in the Brillouin zone. The latter issue leads to results which are not easily comparable with the Dirac viscosity calculation in the continuum limit. We have performed cursory calculations on such a system, but we will  leave the discussion of lattice viscosity calculations when the low-energy states are near generic points in the Brillouin zone to future work. Instead we will consider a simpler lattice model for a Dirac fermion on a square lattice.
When the metric deformation of Eq.~\eqref{eq:cart-metric} is included, the lattice Dirac
model Hamiltonian (on a square lattice in a cylinder geometry periodic in the
$y$ direction, with rational flux $\phi=p/q$ per plaquette) is
\begin{align}
  H = \sum_{n,k_y} &\frac{1}{2}\left(
    ic_{n+1,k_y}^\dag\alpha\sigma^xc_{n,k_y}
    - c_{n+1,k_y}^\dag\sigma^zc_{n,k_y}
    + \text{h.c.}\right) \nonumber \\
  &\quad+ c_{n,k_y}^\dag\sin{(k_y-2\pi\phi n)}\alpha^{-1}\sigma^yc_{n,k_y}
    \nonumber \\
  &\quad+ c_{n,k_y}^\dag\left[2 - m - \cos{(k_y-2\pi\phi n)}\right]\sigma^zc_{n,k_y},
  \label{eq:chern-hamiltonian}
\end{align}
where $c_{n,k_y}$ is a two-component annihilation operator.  This model has a single gapless Dirac cone when $m=0$ or $m=4.$ For $m=0$ ($m=4$) the Dirac cone is located near ${\bf{k}}=(0,0) $ (${\bf{k}}=(\pi,\pi))$ in the Brillouin zone. Like the Hofstadter model, since we have included a magnetic field, the
system geometry is chosen to preserve the magnetic translation symmetry with an integer number of magnetic cells that are $q$ sites wide in the $x$ direction. Since we are using a cylinder geometry the
lattice should have $N_x=lq-1$ so that the boundaries are
commensurate.~\cite{hatsugai_edge_1993}  The Landau levels at $m=0$ and $m=4$ are shown in
Fig~\ref{fig:chern-spectrum}. We immediately recognize the similarities in the two cases, but should point out a major difference, i.e. that the edge states for $m=0$ ($m=4$) are located near $k_y=0$ ($k_y=\pi$). Below we will discuss how this difference affects the results for these two cases. 

\begin{figure}
  \includegraphics{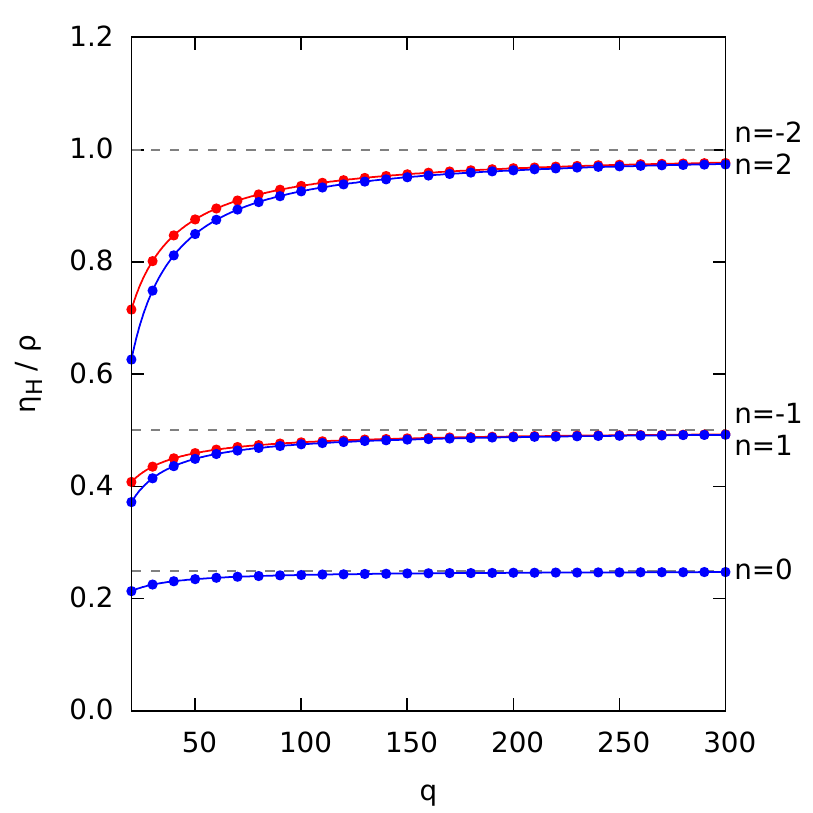}
  \caption{\label{fig:chern-viscosity-transport-1}
    The Hall viscosity of the lattice Dirac
    Hamiltonian~\eqref{eq:chern-hamiltonian}, calculated by
    Eq.~\eqref{eq:viscosity-momentum-transport} with $N_y=51$
    and $N_x=2q-1$. The indicated individual Landau levels are
    filled. The dotted grey line indicates the Hall viscosity of the
    continuum Dirac Landau level given by
    Eq.~\eqref{eq:iqhe-dirac-viscosity}.
  }
\end{figure}

\begin{figure}
  \includegraphics{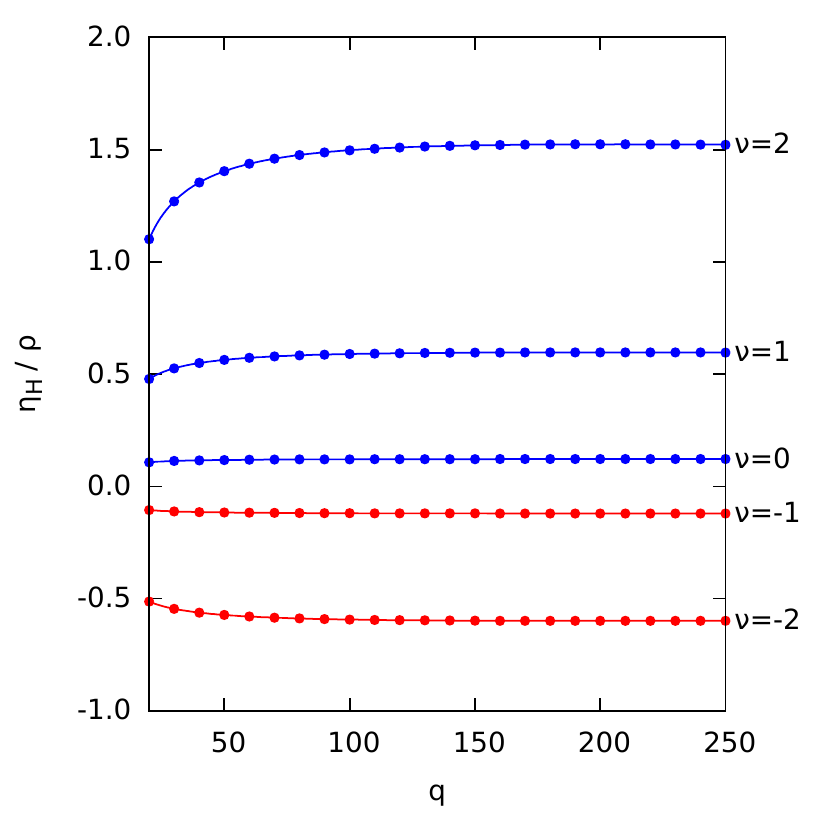}
  \caption{\label{fig:chern-viscosity-transport-thru}
    The Hall viscosity of the lattice Dirac
    Hamiltonian~\eqref{eq:chern-hamiltonian}, calculated by
    Eq.~\eqref{eq:viscosity-momentum-transport} with $N_y=51$
    and $N_x=2q-1$. The Landau levels are filled from the bottom of
    the spectrum through the indicated level. Because filling this way
    causes the number of filled Landau levels to vary with $q$, a
    linear term $0.011q$ has been subtracted from each series.
  }
\end{figure}

The viscosity can again be calculated by projecting the total momentum
onto the right half of the cylinder, as in
Eq.~\eqref{eq:viscosity-momentum-transport}, and then differentiating
with respect to $\alpha^2$. The momentum projection is
\begin{equation*}
  \left\langle P_y \mathcal{P}_R \right\rangle
  = \sum_{k_y} \sum_{j\,\text{occ.}} \hbar k_y
  \left\langle j,\,k_y \right| \mathcal{P}_R \left| j,\,k_y \right\rangle
\end{equation*}
where $\mathcal{P}_R$ projects onto the right half of the
cylinder: 
\begin{equation*}
  \mathcal{P}_R
  = \sum_{x=0}^{\frac{N_x}{2}} \sum_{\sigma=\pm\frac{1}{2}}
  \left|x,\sigma\right\rangle\left\langle{}x,\sigma\right|.
\end{equation*}
The integers $j$ run over the occupied energy eigenstates at a given $k_y.$
For most of the cases we consider we only fill the Landau levels near half-filling. We note that near half-filling the
$n$-th Landau level consists of the states (see Fig. ~\ref{fig:chern-spectrum})
\begin{equation*}
j \in
  \begin{cases}
    \left(N_x + nl ,\, N_x + (n + 1)l\right] & m = 0 \\
    \left(N_x + (n-1)l ,\, N_x + nl\right] & m = 4.
  \end{cases}
\end{equation*}
Notice that the $0$-th Landau level moves from the bottom of the
conduction band at $m=0$ to the top of the valence band at $m=4$. This
is clearly shown  in Fig.~\ref{fig:chern-spectrum}.
\begin{figure}[t]
  \includegraphics{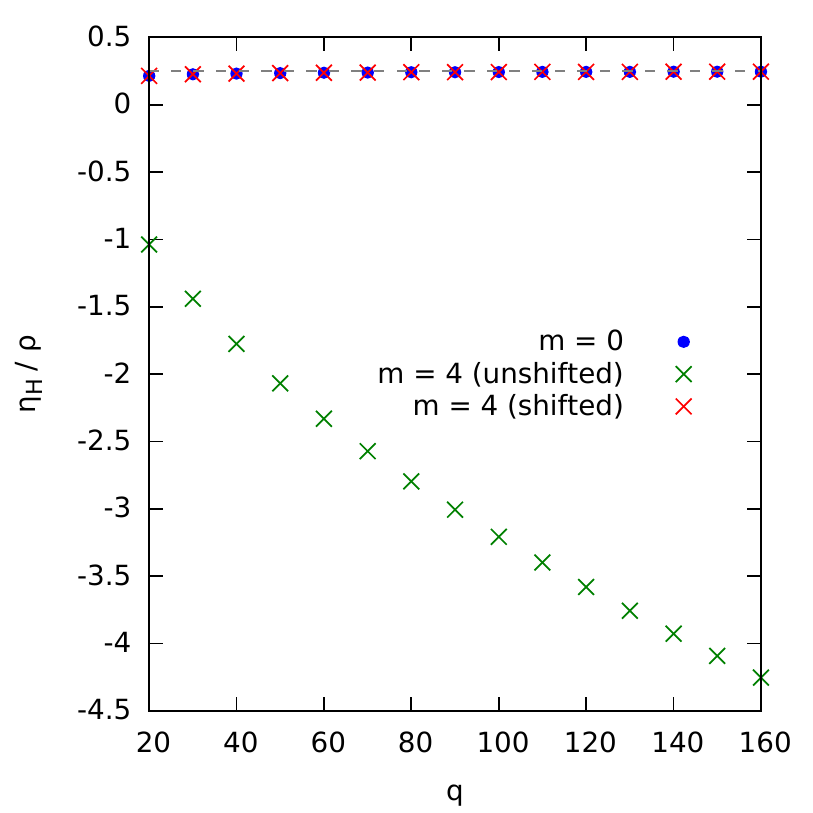}
  \caption{\label{fig:chern-viscosity-transport-0}
    The Hall viscosity of the lattice Dirac model
    Hamiltonian~\eqref{eq:chern-hamiltonian}, calculated by
    Equation~\eqref{eq:viscosity-momentum-transport} with $N_y=51$
    and $N_x=2q-1$. The $n=0$ Landau level Hall viscosity is plotted
    in each case. The $m=4$ Hall viscosity is plotted twice. The
    unshifted plot shows the viscosity when the Brillouin zone is
    unshifted so that the Dirac point is at $k=\pm\pi$. In the shifted
    plot, the Brillouin zone has been shifted so that the Dirac point
    is once again at $k=0$, in which case the Hall viscosity agrees
    exactly with $m=0$.
  }
\end{figure}

The Hall viscosity of the lattice Dirac model was calculated at $m=0$ by the
momentum transport method in
Eq.~\eqref{eq:viscosity-momentum-transport} to obtain the results
in Fig.~\ref{fig:chern-viscosity-transport-1}. The values here represent the viscosity calculations from individually filling (not successively filling) the $n=0,\pm 1,$ and $\pm 2$ Landau levels (where $n=0$ is referenced to the zeroth Landau level of the Dirac point, not the bottom of the entire bandwidth). To help illustrate, we have shown which Landau levels were filled in Fig. ~\ref{fig:chern-spectrum}. We note that the
lattice calculation converges to the continuum value in the large-$q$
(weak magnetic field) limit, i.e. the Hall viscosity of the lattice system
approaches the continuum value in the limit where the magnetic length
$\ell_B$ is much larger than the spacing between unit cells. As the
magnetic field strength increases, so does the effect of the
lattice, with the viscosity taking on $q$-dependent terms:
\begin{align*}
\frac{2\pi\ell_B^2}{\hbar}  \eta_H^{(-2)}
  &\sim 0.9868 + \frac{0.4276}{\sqrt{2\pi}}\frac{a}{\ell_B} + \frac{12.0267}{2\pi}\frac{a^2}{\ell_B^2} \\
\frac{2\pi\ell_B^2}{\hbar}  \eta_H^{(-1)}
  &\sim 0.5018 + \frac{0.0576}{\sqrt{2\pi}}\frac{a}{\ell_B} + \frac{1.7067}{2\pi}\frac{a^2}{\ell_B^2} \\
\frac{2\pi\ell_B^2}{\hbar}  \eta_H^{(0)}
   &\sim 0.2498 + \frac{0.0045}{\sqrt{2\pi}}\frac{a}{\ell_B} + \frac{0.8290}{2\pi}\frac{a^2}{\ell_B^2} \\
\frac{2\pi\ell_B^2}{\hbar}  \eta_H^{(1)}
   &\sim 0.5025 + \frac{0.0857}{\sqrt{2\pi}}\frac{a}{\ell_B} + \frac{1.3075}{2\pi}\frac{a^2}{\ell_B^2} \\
\frac{2\pi\ell_B^2}{\hbar}  \eta_H^{(2)}
   &\sim 0.9802 + \frac{0.6299}{\sqrt{2\pi}}\frac{a}{\ell_B} + \frac{14.1092}{2\pi}\frac{a^2}{\ell_B^2},
\end{align*}
where we have used the relation $qa^2=2\pi{}\ell_B^2$.  As predicted by
the continuum calculation, the Hall viscosity converges to
approximately the same value for positive and negative Landau levels.

Now let us consider what happens if we fill the Landau levels from the absolute bottom of the spectrum, instead
of filling individual Landau levels near the Dirac point as was done in
Fig.~\ref{fig:chern-viscosity-transport-1}. We show the results in Fig.~\ref{fig:chern-viscosity-transport-thru}. Interestingly, we found that the momentum
transported scaled linearly in $q$, and we subtracted off this contribution to make Fig.~\ref{fig:chern-viscosity-transport-thru}. This linear scaling should be expected
because the number of filled Landau levels is proportional to $q$
when filling from the bottom of the spectrum. After subtracting off the
linear term $0.011q,$ as shown in
Fig.~\ref{fig:chern-viscosity-transport-thru}, we find that the
momentum transported indeed saturates to a fixed value in the low-field limit. However, the saturation values do not match the continuum results. Instead,  it is only the 
\emph{differences} in the viscosities between filling the $n$-th and $(n+1)$-th Landau levels that 
exactly matches the continuum value for the viscosity of the added Landau level. This is a surprising result, as it indicates that for lattice systems in magnetic fields the magnitude of the viscosity may be somewhat regularization dependent, but the \emph{difference} in viscosities seems to retain a more universal character. It would be interesting to see if this is a generic feature,  an artifact of this model, or can be attributed to finite size effects.  

Let us now consider the other massless limit of this model  when $m=4.$ We show the viscosity, calculated via momentum transport, in Fig.~\ref{fig:chern-viscosity-transport-0}. The bare result for the viscosity shows a monotonically decreasing function that does not converge for large $q.$ However, to properly interpret this result,  care must be taken to recenter the Brillouin zone. If we keep $m$ fixed, but send
 $k_y\rightarrow{}k_y-\pi$ in the Hamiltonian, then the momentum transport calculation exactly recovers the result at $m=0.$  If the Brillouin zone is not shifted, extra momentum is
transported since the edge states are located near $k_y=\pi,$ and this leads to a different result for the viscosity. We show the $m=0$ case, as well as the shifted and unshifted results for $m=4,$ in Fig.~\ref{fig:chern-viscosity-transport-0}. Generically, when the low-energy Dirac point(s) are away from ${\bf{k}}=0$ in the Brillouin zone there is extra momentum transport due to the overall momentum shift of the cone. While we have been able to adjust the calculation for this simple case (and consequently any case with a single Dirac cone), the question of how to compare the viscosity of multiple Dirac points at generic momenta to the continuum limit remains a topic for future work.

\begin{figure}
  \includegraphics{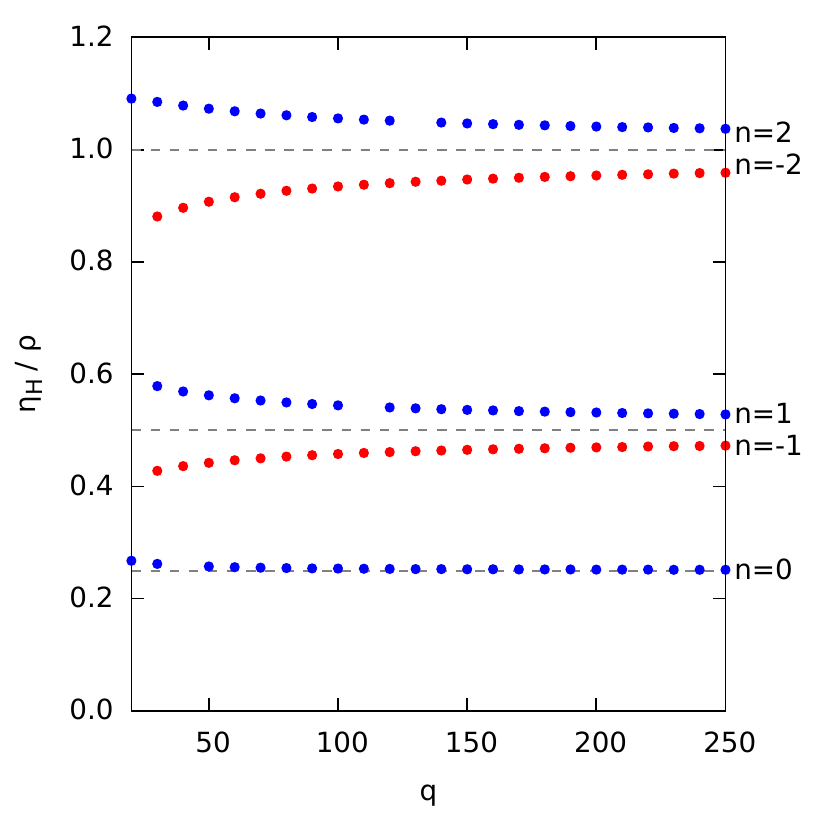}
  \caption{\label{fig:chern-viscosity-polar}
    The Hall viscosity of the lattice Dirac
    Hamiltonian~\eqref{eq:chern-hamiltonian}, calculated by the momentum
    polarization method with $N_x=2q-1$. The indicated individual Landau
    levels are filled. Where points are missing it indicates a failure of
    the fitting required to extract the viscosity.
  }
\end{figure}

The Hall viscosity of the lattice Dirac model was also calculated by the
momentum polarization method as in
Equation~\eqref{eq:viscosity-momentum-polarization}. As in the Hofstadter
model, the correlation function is
\begin{equation*}
  C^{(j)}_{k_y,\alpha}
  = \left\langle j,\,k_y \right| \mathcal{P}_R \left| j,\,k_y \right\rangle.
\end{equation*}
Equation~\eqref{eq:momentum-polarization-corr-func} allows us to
compute $\lambda_{\text{RES}}$ and $\lambda_{\text{OES}}$,
\begin{align}
  \lambda_{\text{RES}}
  &= \prod_{j,k_y} \frac{1}{2}
    \left[\left(1+e^{ik_y\Delta{}y}\right)\right.
    \nonumber \\
  &\qquad + \left.\left(1-e^{ik_y\Delta{}y}\right)
    \left(2\left\langle j,\,k_y \right|\mathcal{P}_R\left|
        j,\,k_y \right\rangle-1\right)\right]
    \nonumber \\
  \lambda_{\text{OES}}
  &= \prod_{j,k_y} \frac{1}{2}
    \left[\left(1+e^{ik_y\Delta{}y}\right)\right.
    \nonumber \\
  &\qquad + \left.\left(1-e^{ik_y\Delta{}y}\right)
    \left(2\Theta(\left\langle j,\,k_y \right|\hat{x}\left|
        j,\,k_y \right\rangle)-1\right)\right].
    \nonumber
\end{align}
Again, $\Theta$ is the Heaviside step function, and $\hat{x}$ is the
$x$-coordinate operator. As with any lattice model, $\Delta{}y$
must be an integer in units of the lattice constant.  The Hall
viscosity obtained this way agrees with the momentum transport
calculation, showing the same convergence to the continuum value of
$\eta_H$ at large $q$ (Figure~\ref{fig:chern-viscosity-polar}). Note
that although both methods show a deviation from the continuum Hall
viscosity at small $q$, the momentum polarization method shows a
smaller deviation and with opposite sign.  Points are missing from these
figures where the fitting required for the momentum polarization
method has failed.

%\begin{figure}
%  \includegraphics{chern-central-charge.pdf}
%  \caption{\label{fig:chern-central-charge}
%    The chiral central charge of the lattice Dirac model edge states, calculated by
%    the momentum polarization method with $N_y=50$
%    and $N_x=2q-1$. The indicated individual Landau levels are
%    filled. Where points are missing it indicates a failure of the
%    momentum polarization method to converge.
%  }
%\end{figure}

\section{Discussion and Conclusion}

We have applied two techniques for calculating the Hall viscosity in
integer quantum Hall systems. Both methods seem to capture similar results for continuum models, and, more interestingly, both were successfully applied to lattice models. Our original momentum transport method
gives results in agreement with the momentum polarization method
previously described, especially in the weak magnetic field limit. While we have seen that there are lattice-scale dependent contributions to both the momentum transport and the momentum polarization at strong magnetic field, these corrections seem to be method dependent, at least for the system sizes and parameters we have chosen.

We have demonstrated that either method can
determine the Hall viscosity of an isotropic system. However,  further
work will be required to fully characterize the Hall viscosity of
anisotropic systems, though we demonstrated that two of the three
non-vanishing Hall viscosity coefficients can already be computed using our methods and geometry. We found that the Hall viscosity coefficients were dependent on the amount of anisotropy, though it
is  unknown if this variation survives in the thermodynamic limit.

Finally, we have calculated the viscosity for the Dirac Landau-level system, and shown that there is a relationship between the Hall viscosity
of continuum Schr\"odinger and Dirac Landau levels. Namely,  that the
latter recovers the viscosity of the former in the infinite mass
($m\rightarrow\infty$) limit.  Our methods reproduce the known results
for these models. Furthermore, we show that both the Hofstadter and
lattice Dirac models approach the appropriate continuum Hall viscosity
as the magnetic field $B\rightarrow{}0$, with deviations at stronger
fields that depend on the lattice scale. Our results for the lattice
Dirac model suggest that it may be the viscosity difference between
Landau level fillings which is actually quantized (in units of density) in lattice
regularized models. We uncovered some difficulties in treating lattice Dirac systems
with multiple Dirac points, when they are located at generic points in the Brillouin zone; further work will be required to treat
some systems of interest, such as graphene. It is also of interest to understand the competition between time reversal breaking arising from the applied magnetic field, and an intrinsic time reversal breaking coming from a massive Dirac model, i.e. a Chern insulator. The latter is also expected to have a non-vanishing, field-independent, contribution to the viscosity\cite{hughes2011}, though it has yet to be calculated in a lattice regularization. The article Ref. \onlinecite{hassan2015} addresses some aspects of this last topic.

\begin{acknowledgements}
We thank S. Ramamurthy and H. Shapourian for discussions. TLH is supported by the US National Science Foundation under grant
DMR 1351895-CAR.
\end{acknowledgements}

\end{document}